\documentclass[reprint, onecolumn,secnumarabic,amssymb,notitlepage,nobibnotes, nofootinbib, 12pt]{revtex4-1}
\usepackage{bm}           
\usepackage{graphicx}
\usepackage{lineno}
\usepackage{amsmath,amssymb,mathrsfs}  
\usepackage[pdftex,bookmarks,colorlinks,breaklinks]{hyperref}  
\usepackage{bm}
\setcitestyle{numbers}
\newcommand{\be}{\begin{equation}}
\newcommand{\ee}{\end{equation}}
\newcommand{\ber}{\begin{eqnarray}}
\newcommand{\ear}{\end{eqnarray}}
\newcommand{\bs}{\begin{subequations}}
\newcommand{\es}{\end{subequations}}
\newcommand{\bc}{\begin{center}}
\newcommand{\ec}{\end{center}}
\newcommand{\ba}{\begin{array}}
\newcommand{\ea}{\end{array}}

\newcommand{\sfrac}[2]{{\textstyle\frac{#1}{#2}}}
\renewcommand{\:}[2]{{\textstyle\frac{#1}{#2}}}

\newcommand{\forget}[1]{\iffalse#1\fi}
\newcommand{\forgetmenot}[1]{\iftrue#1\fi}
\newcommand{\del}{\nabla}

\renewcommand{\div}{\hskip0.9pt{\mathsf{div}\hskip2pt}}

\newcommand{\curl}{\hskip0.9pt{\mathsf{curl}\hskip2pt}}

\newcommand{\Qt}{Q^{(T)}}
\newcommand{\Qtb}{{\bar Q}^{(T)}}

\newcommand{\nn}{\nonumber}

\newcommand{\fr}{\begin{flushright}}
\newcommand{\efr}{\end{flushright}}
\newcommand{\fl}{\begin{flushleft}}
\newcommand{\efl}{\end{flushleft}}

\def\case#1/#2{\textstyle\frac{#1}{#2} }

\renewcommand{\div}{\mathsf{div}}

\newcommand{\cavK}{\mathcal{K}}
\newcommand{\curlA}{\hskip0.9pt{\mathsf{\bar{\curl}}\hskip2pt}}
\newcommand{\disA}{\hskip0.9pt{\vec{\mathsf{dis}\hskip2pt}}}
\newcommand{\sdel}{\vec{\nabla}}
\newcommand{\V}{_{\hskip-0.8pt\mathrel{\vcenter{\hbox{\tiny\ooalign{\raise 1.5pt\hbox{\textsf{V}}}}}}}}
\newcommand{\T}{_{\hskip-0.8pt\mathrel{\vcenter{\hbox{\tiny\ooalign{\raise 1.5pt\hbox{\textsf{T}}}}}}}}

\begin{document}
\title{Second order perturbation theory: A covariant approach involving a barotropic equation of state}
\author{Bob Osano}
\email{bob.osano@uct.ac.za} 
\affiliation{Cosmology and Gravity Group, Department of Mathematics and Applied Mathematics, University of Cape Town \\}
\affiliation{Academic Development Programme, Science, Centre for Higher Education Development, University of Cape Town, Rondebosch 7701, Cape Town, South Africa}
\begin{abstract}
We present a covariant and gauge-invariant formalism suited to the study of \emph{second-order effects} associated with higher order tensor perturbations. The analytical method we have developed enables us to characterize pure second-order tensor perturbations about FLRW model having different kinds of equations of state. Our analysis of the radiation case suggests that it may be feasible to examine the CMB polarization arising from higher order perturbations. 
\end{abstract}
\pacs{}
\date{\today}
\maketitle
\section{Introduction}
The study of cosmology is now firmly data driven thanks to the availability of large amount of high quality data from numerous large-scale surveys such as those from galaxies red-shifts, the measurements of the CMB temperature anisotropies and polarization~\cite{TCM07}. The standard model of cosmology is parametrized by six values and is based on a flat universe that is dominated by a cosmological constant ($\Lambda$) and cold dark matter (CDM), with initial Gaussian distribution, and inflation-seeded adiabatic fluctuations. Although this model, broadly speaking, successfully describes all existing CMB data \cite{WMAPNine, Bennett}, the standard model raises questions among which are, what is the physics of inflation? Are the initial fluctuations adiabatic? The finer features of the data also raises questions on the veracity of the features of the underlying model. These questions demand a refinement or reexamination of the theoretical and data analysis considerations. To this end, it is worth considering the role that of nonlinear perturbations might have on some of the measurements that yield the 6 parameters. 
 
\par Studies of small fluctuations that are thought to have given rise to large-scale temperature anisotropies and polarization of the CMB are usually treated with first-order relativistic perturbation theory ~\cite{KODSA84, MUK92, Bardeen:1980kt, DUR04, MMB98}. Second-order perturbation theory is increasingly becoming necessary when probing scales where linear theory becomes in accurate. This is particularly important because there is no clear method within the linear theory that can help determine when the perturbations have become too large for the theory to handle ~\cite{Nakamura:2006rk, Clarkson:2003af} and hence the need for a second order relativistic perturbation theory.  
For these reasons, second order perturbation theory has received considerable attention ~\cite{Acquaviva:2002ud, Nakamura:2006rk,Clarkson:2003af, Brand07, Osano:2006ew, Teresa:2006af,Mena:2002wq,Matarrese:1997ay,Finelli:2006wk,Noh:2004bc,Clarkson:2003af,Langlois:2005qp, Langlois:2005ii,Ananda:2006af,Nakamura:2006rk, Malik2003mv}.
\par Second-order cosmological perturbations dates back to ~\cite{Tomita67}, who extended Lifshitz's linearized theory. In that work, the author demonstrated that the second-order density contrast led to increased first-order density contrast over time when the perturbation was not too large. It was shown that at second order gravitational waves could be induced by deformed density perturbations even where first-order perturbations were non tensorial. Recently, the author of \cite{Clarkson:2003af} examined a similar effect in the 1+3 formalism. Second order perturbation theory has been used as primary tool for considering nonlinear dynamics. For example, the authors of ~\cite{Mena:2002wq} looked at second-order perturbations of a flat dust FLRW models with a cosmological constant. These authors considered the evolution of second order perturbations in flat FLRW models with $\Lambda\neq 0$ and having a dust equations of state. They showed that these perturbations tend to be constants in time, in agreement with the cosmic no-hair conjecture. The authors of ~\cite{Matarrese:1997ay} considered the case without a cosmological constant, where gauge transformations at second order were introduced and used to study the gauge dependency of perturbations. Second order effects during inflation were studied in ~\cite{Acquaviva:2002ud}, where the prediction of the bispectrum of perturbation from inflation was examined. The full relativistic treatment of second order perturbation theory has also been considered in ~\cite{PC96, MM97, MMB98, Noh:2004bc}. In \cite{PC96}, the behavior of light rays in perturbed FLRW models is studied and the redshift between an observer and the surface of last scattering to second order in the metric perturbation is explicitly calculated. Here in we find the problem associated with the size of perturbations explicitly state. In particular, it is pointed out that there is no guarantee that second order effects are significantly smaller than those at first order, given that the large length scales associated with the problem could give rise to large pre-factors. The implications of this on the linear theory are significant and need to be examined, something that we partly do in this article. Meanwhile, second order curvature perturbations on super-Hubble scales after inflation were studied in \cite{Malik2003mv, BCLM04}. On the other hand, the authors of ~\cite{Bartolo:2006cu, Bartolo:2006fj} have shown that second order effects may lead to detectable non-Gaussianity in the CMB, while those of ~\cite{Mollerach:2003nq, BMMR} have considered second order contributions to CMB polarization in the metric approach. The authors of \cite{BMMR} studied the B-mode CMB polarization where they found that such contributions make up part of a contamination in the detection of the primordial tensor modes if the tensor to scalar ratio $r$ is smaller than a few $\times10^{-5}$.

The literature cited above, with a few exceptions, develop and apply metric based formalisms in their analysis of higher-order perturbations. We are interested in a 1+3 covariant and gauge invariant approach to complement these efforts. In ~\cite{Clarkson:2003af}, a covariant approach to nonlinear perturbation theory was initiated, and the formalism used to study second order gravitational waves sourced by first order density perturbations and second order density perturbations sourced by gravitational waves at first order. However, only the dust equation of state was considered in that work. We extend this formalism to barotropic fluids and use the findings to argue for a second order effect on the CMB. This paper is organized as follows: In section (\ref{sec1}) we give the background to the 1+3 covariant and gauge Invariant approach. Section (\ref{sec2}) discusses perturbation theory in the 1+3 formalism. The barotropic fluid is considered in section (\ref{Bar}).  Section (\ref{dusty}) gives the analysis of the system with dust equation of state, complete with both analytical and numerical solutions. Section (\ref{Bar}) then gives the analysis of the case with barotropic equation of state. Here too, both the analytical and numerical solutions are given. Section ({\ref{CMB}) looks at second order effects on CMB, while the conclusion and future work is presented in section (\ref{Conc}).

\section{\label{sec1}1+3 formalism}
All variable are defined on a model that has a FLRW geometry of curvature $\cavK$. This geometry is intrinsically linked to the 4-velocity, $u^a$, given by the vector tangent to the fundamental observer world-lines such that $u^{a}=dx^{a}/d\tau$ and $u^{a}u_{a}=-1$. Based on the 4-velocity, we can defined local variables that characterize the model. To achieve this, we need two operators, (i) the projection tensor $h_{ab} (=g_{ab}+u_a u_b)$ which projects into the tangent 3-spaces that are orthogonal to $u^{a}$ for the case where vorticity vanishes. (ii) We also need a covariant derivative $D_{a}\equiv( )_{;a}$. In general, the first covariant derivative of the 4-velocity $u_{a}$ is given by\begin{eqnarray}D_{b}u_{a}=\omega_{ab}+\sigma_{ab}+\frac{1}{3}\Theta h_{ab} -A_{a}u_{b},\end{eqnarray} where $\Theta=D^{a}u_{a}$ is the expansion\footnote{ Let $\mathscr{L}$ be a typical mean distance of some fluid behavior. From a typical volume $\mathscr{L}^3$, it follows that the rate of change of volume is $3\dot{\mathscr{L}}\mathscr{L}^{2}$. Expansion is defined as the rate of change normalized by volume. In particular; $3\dot{\mathscr{L}}/{\mathscr{L}}=\Theta$. }, $\sigma_{ab} =D_{(a}u_{b)} $ is the shear tensor, $\omega_{ab}=D_{[a}u_{b]} $ is the vorticity and $A_{a}=\dot{u}_{a}=u_{b}\nabla^{b}u_{a}$ is the relativistic acceleration vector ( we note that the spatial derivative is not projected in this case). It indicates the extent to which matter can be moved by none-gravitational and none-inertial forces.
A perfect fluid filled FLRW background is characterized by $\sigma_{ab}=\omega_{ab}=u^{a}=0.$ The non-zero scalars in this background are the expansion parameter ($\Theta$), the 3-Ricci curvature $^{(3)} R (=\cavK)$, the energy density $\mu(=T_{ab}u^{a}u^{b})$ and isotropic pressure $p (=\frac{1}{3}T_{ab}h^{ab}),$ where $T_{ab}$ is the energy-momentum tensor. A detailed account of the 1+3 formalism may be found in \cite{EMM,EB89}. Although only the special case of a perfect fluid  filled model is considered in this article, the formalism can be extended to other types of equation of states.

\section{\label{sec2}Perturbation theory in the 1+3 formalism}
The framework for perturbation theory, in the 1+3 formalism, was developed in \cite{EB89}. The formalism can be considered a {\it top-down} approach in the sense that one begins with the big picture and then breaks it down into smaller parts. To be more precise, the methodology requires one to begin with propagation and constraint equations for a fully perturbed model then {\it linearize} about a background of choice. It is standard, and we do the same in this analysis, to choose the FLRW model for a background. This is because our real universe is, at least on large scale, well described by this model. The 1+3 approach to perturbation theory is different to the standard gauge-invariant metric perturbation theory where one begins with variables representing a given background and then perturbs them to desired order before finding the equations of motion for perturbed model. This approach may be thought of as a {\it bottom-up} approach in the sense that one pieces together separate parts to generate more complex systems and thereby rendering the original parts constituents of the emergent system. Our intention is not to compare the two approaches but to provide a basic framework against which the work presented in this article can be understood. Imagine that there exist several special {\it filters} and that these filters are able to separate gauge-invariant perturbations according to their sizes (or {\it order}). It should be emphasized that these filters are not physical but conceptual. Of course the first question that needs addressing is how one might go about constructing them. The first-order filters are constructed using conditions outlined by the {\it Stewart and Walker} lemma \cite{Stewart:1974uz}. This lemma requires that a quantity vanish in the background, be a constant scalar field or a linear combination of products of delta functions for it to be gauge-invariant. One can extend this to define {\it second-order-filters}. In particular one requires the filter to filter-out gauge-invariant quantities that are not zeroth or first-order \cite{Bruni:1992dg}. FIG (\ref{sieve}) gives a schematic presentation of the approach. 
\begin{figure}[htbp]
\begin{center}
   \includegraphics[width=6in]{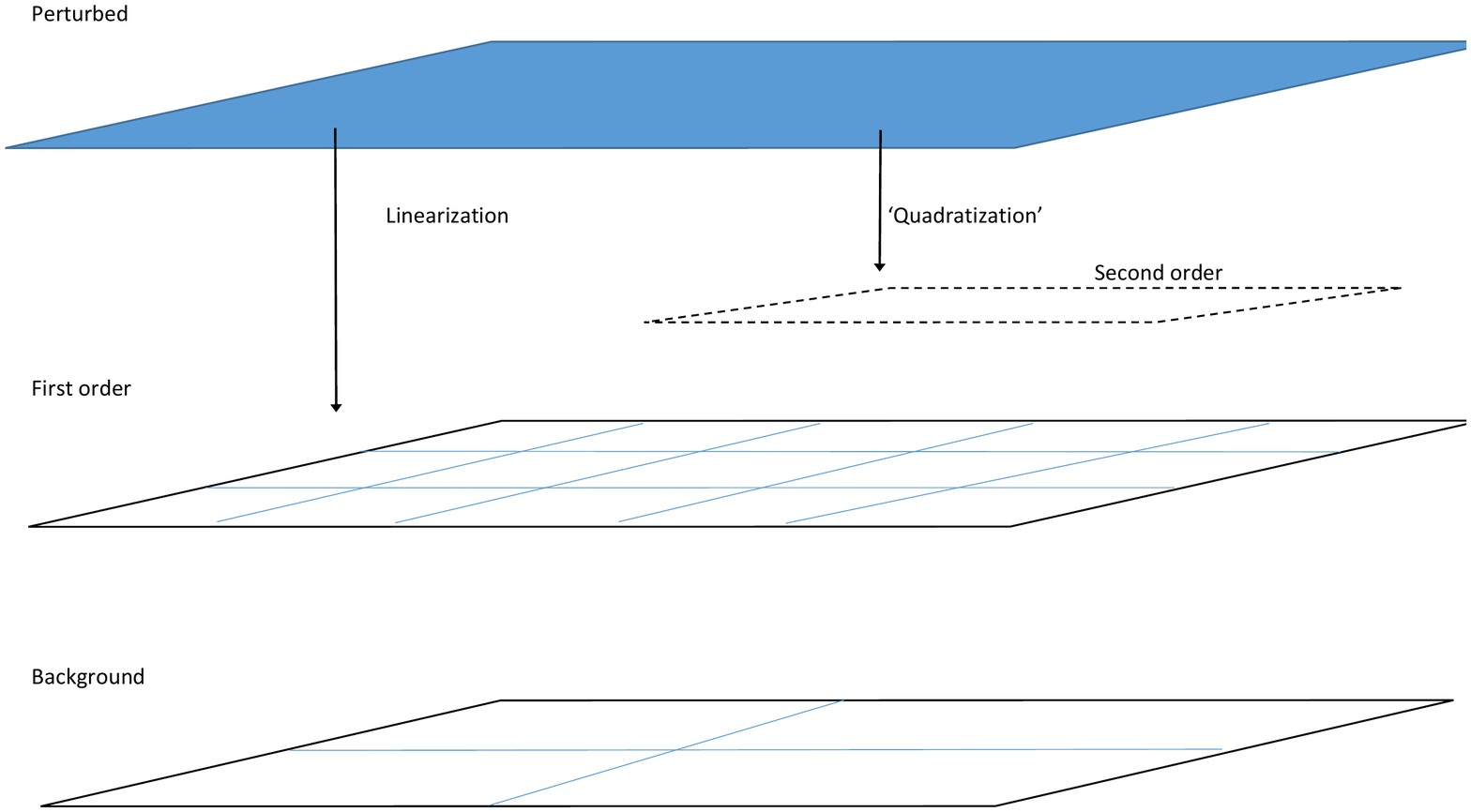}
\caption{\label{sieve}}
\end{center}
\end{figure}
There is an inherent difficulty though in this formalism, one that is not dissimilar to that in the metric approach, which has to do with how to decide when a particular order of perturbation is `too big' and what to do with such. 

In the present article, we will deal with an enhancement of 1+ 3 approach to perturbations to allow for the analysis of perturbations that are of the second order type. We coin the word {\it Quadratization} to denote the act of dropping terms of order higher than 2 in our perturbation scheme. This would still leave a coupling of first and second order perturbations and require a means of separating them if we wish to analyze uniquely second order effects. We will describe in the next section how this may be done.

\subsection{First Order Dynamics}
First order perturbations about a FLRW background are relatively easy to characterize. The first batch are the covariant derivatives $\mathcal{X}_{a}=D_a\mu$, $\mathcal{Z}_a=D_a\Theta$,  and $ \mathcal{C}_a=a^3 D ^{(3)}R$, and are \emph{first-order gauge-invariant} (FOGI) variables corresponding respectively to the spatial fluctuations in the energy density, expansion rate and spatial curvature of a fluid flow. It is worth emphasizing that these quantities are FOGI because they are co-variantly defined and that they vanish exactly in the background FLRW spacetime, in line with the perturbation scheme outlined in the previous section.
 The second batch of FOGI are the shear tensor $\sigma_{ab}=D_{\langle a}u_{b\rangle}$, the electric part of the Weyl tensor $E_{ab} (\equiv C_{cdgh}{C^{gh}}_{ef}{h^{c}}_{a}u^{d}{h^{e}}_{b}u^{f}$ where $C_{cdef}$ is the Weyl tensor) and the magnetic part of the Weyl tensor $H_{ab}(\equiv \frac{1}{2}\epsilon_{cdef}{h^{c}}_{a}u^{d}{h^{e}}_{b}u^{f}){\color{red}C}$. Together, these quantities represent effects on either geometry or matter. We assume that the interaction between the geometry and matter is given by Einstein gravitational field equations (EFE): \[R_{ab}-\frac{1}{2}Rg_{ab}=T_{ab},\] where $R_{ab}$ is the Ricci tensor, $R$ is the Ricci scalar and $g_{ab}$ the metric. Note that we have taken the cosmological constant $ \Lambda=0$. Three sets of equation emanate from the EFE and the integrability conditions associated with it. From the {\it Ricci identities} associated with the vector field $u^{a}$ ($2\nabla_{[a}\nabla_{b]}u^{c}={{R_{ab}}^{c}}_{d}u^{d}$), one obtains the {Raychaudhuri equation,} the vorticity propagation equation, the shear propagation equations, the shear divergence constraint, vorticity divergence identity constraint and the magnetic Weyl tensor constraint. Of these, only the Raychaudhuri equation (\ref{Ray}) tells us something about the background dynamics.
 
 \begin{eqnarray}\label{Ray}\dot{\Theta}=-\frac{1}{3}\Theta^2-\frac{1}{2}(\mu+3p)-2\sigma^2+2\omega^2+D_{a}A^{a}+A^2,\end{eqnarray} 
 where $\sigma^{2}\equiv\sigma_{ab}\sigma^{ab} $, $\omega^{2}\equiv\omega_{ab}\omega^{ab}$ and $A^2\equiv A_{a}A^{a}$. Since the definition of orthogonal projection onto hyper-surfaces demand that $\omega_{ab}=0$, it follows that $\omega^{2}=\sigma_{ab}\sigma^{ab}=0.$ Note that the {\it background-filter} renders $\sigma^2=D_{a}A^{a}=A^2 =0$ leaving \begin{eqnarray}\label{thetadot}\dot{\Theta}=-\frac{1}{3}\Theta^2-\frac{1}{2}(\mu+3p).\end{eqnarray} This is because these terms are not {\it zeroth-order} as required of background terms. It is also clear that a {\it first-order-filter} will not capture $\sigma^2$ or $A^2$ as they are products of first-order variables. The {\it second-order-filter} will nevertheless catch $D_{a}A^{a}(\neq 0)$. In theory, this means that the different orders of perturbations could contribute to the rate of expansion in the fully nonlinear set up, however the contributions from these terms are often insignificant compared to those of zeroth-order terms and are usually neglected. This has however become a subject of debate in recent times \cite{Buchert2015,Green2014}. We assume that such back-reactions on zeroth-order dynamics are negligible. The twice-contracted Bianchi identities guarantee that the total energy-momentum is conserved. The energy conservation equation gives us the other background propagation equation, namely 
 \begin{eqnarray}\label{mudot}\dot{\mu}=-(\mu+p)\Theta.\end{eqnarray} In a general mathematical setup, we would also require the equation for $\dot{p}$ but physical equation of state considerations for what we would like to study demand that either $p=0$ ( as is the case of {\it dust} equation of state) or $p=$w$\mu$ ( as is the case for {\it barotropic} equation of state). These two options imply that $\dot{p}$ is already accounted for in the propagation equation for ${\mu}$.
 
Let us now consider the consequence of {\it first-order-filters} on the general propagation equations for first order scalars, vectors and tensors. As mentioned above, the first order {\it scalar-vector-tensor} (SVT) quantities with respect to the FLRW background are $\sigma_{ab}, E_{ab}$ and $H_{ab}$. We often think of these as tensors given their transformation properties. However, with the exception of $H_{ab}$, the rest may be split into scalar, vector and tensor parts. The $H_{ab}$ will have vector and tensor parts, as we will describe below. We use term $SVT$ ({\it scalar-vector-tensor} ) to refer to such quantities. It is easy to obtain the propagation equation for each of these quantities. In this section, we present them in a coupled system given by the matrix:
\ber\label{MAT1}
\left[
\begin{array}{ccc}
\label{prop1}\sigma_{ab} \\
E_{ab} \\
H_{ab}
\end{array}
\right]^{.}=
\left[
\begin{array}{ccc}
 -\frac{2}{3}\Theta &  -1 & 0  \\
  -\frac{1}{2}(\mu+ p)& -\Theta  & \curl  \\
  0 & - \curl &  -\Theta 
\end{array}
\right]\left[
\begin{array}{ccc}
\sigma_{ab} \\
E_{ab} \\
H_{ab}
\end{array}
\right]
+\left[\begin{array}{ccc}D_{\langle a}A_{b\rangle}\\0\\0\end{array}\right]
,\end{eqnarray} where $\curl \sigma_{ab}\equiv\varepsilon_{ef\langle a}D^{e}{\sigma_{b\rangle}}^{c}$ (the permutation tensor $\varepsilon_{abc}=u^{d}\eta_{abcd}$ is a volume element \cite{EHvE98}. The shear propagation equation is obtained from {Ricci identities}, while the propagation equations for $E_{ab}$ and $H_{ab}$ come from the Bianchi identities $\nabla_{[a}R_{bc]de}=0$, where the Riemann tensor $R_{abcd}$ be split into the Ricci tensor $R_{ab}$ and the Weyl curvature tensor $C_{abcd}$. The 1 + 3 splitting of these quantities, and EFE, together with the once-contracted Bianchi identities give the propagation equations for $E_{ab}$ and $H_{ab}$ \cite{EllisCar}. We note that the full nonlinear set of equations have been subjected to a {\it first-order-filter} to yield matrix  (\ref{MAT1}). Unlike the traditional matrices where the entries are either scalar or complex quantities, the square matrix in this case is an operator owing to the presence of  the $\curl$ terms. However, a Fourier decomposition of the system would yield a simple square matrix with time depended variables. Similarly, the propagation equations for first-order {\it scalar-vector} (SV) quantities are \begin{eqnarray}
\left[
\begin{array}{c}
\mathcal{X}_{a}\\  
 \mathcal{Z}_{a} 
\end{array}
\right]^{.}=
\left[\begin{array}{cc} -\frac{4}{3}\Theta &~~~\mu\\-\frac{1}{2}&~~~   \Theta   \end{array}\right]
\left[
\begin{array}{c}
\mathcal{X}_{a}\\  
\mathcal{Z}_{a} 
\end{array}
\right]. \end{eqnarray} These are derived from the time derivative of $\mathcal{X}_{a}$ and $\mathcal{Z}_{a}$, the commutation relation (\ref{app:2}) and Eqs.(\ref{thetadot}) and (\ref{mudot}). One also obtains the following FOGI constraints from the above mentioned Ricci and Bianchi identities.
\ber\div(\sigma_{a},~~ E_{a},~~H_{a})&=&(\frac{2}{3}\mathcal{Z}_{a},~~\frac{1}{3}\mathcal{X}_{a},~~0)\label{const1}\\
\curl (\sigma_{ab},~~\mathcal{X}_{a})&=&(H_{ab}, ~~0)\\ \label{presCon}D_{a}p&=&-(\mu+p)\dot{u}_{a},\ear
where $\mathsf{div}(\sigma_{a})\equiv
D_{e}{\sigma_{a}}^{e}$. We note that constraint (\ref{presCon}) is redundant if either dust or barotropic equations of state are assumed, as we will do in this article.
These sets of propagation and constraint equations allow the study of linear order dynamics, something that has been done extensively.
\subsection{\label{HO}High-order contributions}
Our perturbation scheme relies on reducing full nonlinear equations to a set of propagation and constraint equations that is interpreted as perturbations about a chosen background. We saw in the previous section how {\it linearization} led to the first order perturbations about the FLRW. In particular, we began with full propagation equations for each quantity and proceeded to drop all terms that were made of products first order quantities. The principle here is that these products are of order higher than first order and are too small to play a role at this level. The interpretation is not so simple when we begin with propagation equations and drop terms that are higher than second-order (i.e {\it Quadratize}). This ensures that we only have up to second-order terms. The subtle implication here is a back-reaction of second order quantities on first-order dynamics. This is captured in the following system:
\begin{eqnarray}\label{Alldot}
\left[
	\begin{array}{ccc}
		\sigma_{ab} \\
		E_{ab} \\
		H_{ab}
		\end{array}
\right]^{.}=
\left[
	\begin{array}{ccc}
 		-\frac{2}{3}\Theta &  -1 & 0  \\
  		-\frac{1}{2}(\mu+p)& -\Theta  & \curl  \\
  		0 &  -\curl &  -\Theta 
	\end{array}
\right]
\left[
	\begin{array}{ccc}
		\sigma_{ab} \\
		E_{ab} \\
		H_{ab}
	\end{array}
\right]
+\left[
	\begin{array}{ccc}
	-\sigma_{c\langle a}{\sigma_{b\rangle}}^{c}+~D_{\langle a}A_{b\rangle}+A_{\langle a}A_{b\rangle}\\+3\sigma_{c\langle a}{E_{b\rangle}}^{c}+2\epsilon_{cd\langle a}A^{c}{H_{b\rangle}}^{d} \\
	+3\sigma_{c\langle a}{H_{b\rangle}}^{c}-2\epsilon_{cd\langle a}A^{c}{E_{b\rangle}}^{d}
	\end{array}
\right]
\end{eqnarray} The obvious additions are the quadratic products  in the last column matrix that clearly represent second order contribution to how FOGI variables evolve. These products are gauge invariant \cite{Bruni:1996im}. The spatial fluctuation quantities satisfy the system:
\begin{eqnarray}
\left[
\begin{array}{c}
\mathcal{X}_{a}\\  
 \mathcal{Z}_{a} \\
 A_{a}
\end{array}
\right]^{.}=\left[
\begin{array}{ccc}
 -\frac{4}{3}\Theta &~~~ -(\mu+p) &- (\mu+p)\Theta   \\
 -\frac{1}{2} (\mu+3p)&~~~  - \Theta   & \frac{1}{3}\Theta^2+\frac{1}{2}(\mu+3p) \\
-\frac{4}{3}\Theta \mathcal{I}&~~~ -(\mu+p)\mathcal{I}&- (\mu+p)\mathcal{I}\Theta  
\end{array}
\right]
\left[
\begin{array}{c}
 \mathcal{X}_{a}\\  
 \mathcal{Z}_{a} \\A_{a}
\end{array}
\right] - \left[
\begin{array}{c}
{\sigma_{a}}^{b} \mathcal{X}_{b}\\  
 {\sigma_{a}}^{b}\mathcal{Z}_{b} \\\mathcal{I}{\sigma_{a}}^{b}X_{b}
\end{array}
\right]\end{eqnarray}
where $\mathcal{I}=-p/(\mu+p)$. Again the column matrix demonstrates the existence of order coupling, as is evident by the presence of the first order shear tensor that couples to spatial gradients of background scalars. This coupling, although interesting, increases the level of difficulty in dealing with the system. The constraint equation to the above system are given by:
\ber 
\div(\sigma_{a}, E_{a})&=&(\frac{2}{3}\mathcal{Z}_{a},~~\frac{1}{3}\mathcal{X}_{a}+\epsilon_{abc}\sigma^{bd}{H^{c}}_{d}) \\ \div(H_{a})&=&( -\epsilon_{abc}\sigma^{bd}{E^{c}}_{d})\\
\curl(\sigma_{ab},~~\mathcal{X}_{a})&=&(H_{ab}, ~~0)\label{H1}
\ear
 Unlike in first-order dynamics where the different types of perturbations about FLRW (with p=0) decouple, the different types of perturbations are coupled in second-order dynamics. These coupling present several challenges. The two technical challenges encountered in studies of perturbations beyond the first order are (1) order mixing i.e. {\it first - second} order perturbations mixtures, and (2) one type of perturbation sourcing another type. These are in addition to problems associated with gauge issues which would arise were the variables not defined as gauge invariant from the beginning. We would of course want to know what residual gauge related issues show up at second, something that has been discussed in \cite{Bruni:1996im}. One way of skirting around these some of these problems is to develop methods for extracting second order gauge invariant quantities such as the ones in \cite{CB2011} and which we outline in the next section. We only present second-order {\it pure tensor extractor} as an illustration. Other extractors can be found in \cite{CB2011}.

\subsection{\label{ExtraK} Higher order perturbation-type extraction} It is a well known fact that propagation equations for scalar, vector and tensor perturbations couple when one considers nonlinear perturbations. One would require a method for extracting one type of perturbations, say tensorial type, if the objective is to analyze the role that such perturbations plays in higher order dynamics. This has been a long standing challenge and one of the reasons why higher order perturbations have not received greater attention, if one were to compare with the attention given to linear perturbations. There of course other reasons such as the difficulty of linking higher order perturbations to cosmological observations on the one hand and the belief that linear order perturbations was sufficient for the link between theory and observations. These sets of propagation and constraint equations allow the study of linear order dynamics. We now know that with increased precision in cosmological measurements and data analysis, new and hitherto unexplained features are becoming apparent thereby necessitating the development of high order perturbation theory to try to explain what linear order perturbations schemes are unable to. A comprehensive scheme for extracting higher order perturbations in the 1+3 formalism was first given \cite{Clarkson:2011td} and is the scheme that we will use. The reader is referred to that article for the full explanation, we nevertheless present an overview here.
 
 The scheme is based on the assumption that all variables are defined on a background that has FLRW geometry of curvature $\cavK$. It can be shown that all relations developed below hold for objects of any perturbative order $m$. All $VT$ and $SVT$ ( rank-1 and -2 tensors ) given here are defined to be orthogonal to $u^a$. $SVT$ tensors are symmetric and trace-free (PSTF) and following \cite{Maartens:1996ch}, we use angle brackets on indices as a reminder of this fact. Following \cite{Clarkson:2011td}, we define a conformal spatial covariant derivative acting on scalars or spatial tensors as $\sdel_c=\mathscr{L} D_{c}$, where $\mathscr{L}$ is the scale factor and $D_c$ is the 1+ 3 spatial derivative defined above. The operator $\sdel_a$ has the property that it commutes with the time derivative operator $u^a\del_a$ (i.e $D_{c}[u^a\del_a Y_{ab}]\equiv u^a\del_a[D_{c} Y_{ab}]$) for any SVT (rank-2 tensor) $Y_{ab}$. 
 
This operator allows for the definition of the irreducible parts of the spatial derivative of any SVT tensors, in particular the divergence, the curl and the distortion which are respectfully given by:\ber\label{div}\div {Y}_{b\dots c}&=& \sdel^a Y_{ab\dots c}\\\ \label{curl} \curlA Y_{ab\dots c}&=&\varepsilon_{de\langle  a}\sdel^d Y_{b\dots c\rangle}^{~~~~~e}\\ \label{dis} \disA Y_{ca\dots b}&=&\sdel_{\langle c}Y_{a\dots b\rangle}. \ear Any SVT tensor may then be decomposed as follows: \ber\label{SplitT} Y_{ab}&=&\mathcal{S}_{ab}+\mathcal{V}_{ab}+\mathcal{T}_{ab}\nonumber\\ &=&\sdel_{\langle a}\sdel_{b\rangle}\mathcal{S}+\sdel_{\langle a}\mathcal{V}_{b\rangle}+\mathcal{T}_{ab}, \ear where the non-local scalar part is curl-free ($\curl\mathcal{S}_{ab}=0$), the vector part is solenoidal ($\div\mathcal{V}=0\Rightarrow\div\div \mathcal{V}=0$ and $\div \mathcal{V}_a\neq0$), while the tensor part is transverse, $\div \mathcal{T}_{a}=0$. The question of how to form \emph{local} scalar, vector and tensor quantities from $Y_{ab}$ and relate them to the non-local split given above was answered in \cite{Clarkson:2011td}. It was also shown that when $Y_{ab}$ obeys a wave equation in the form $\mathcal{L}[Y_{ab}]=\mathcal{W}_{ab}$, where $\mathcal{W}_{ab}$ is the source and $\cal L$ contains time derivatives and Laplacians, and any derivative operations which preserve the rank of $Y_{ab}$~-- i.e., $\curlA$, $\disA\div$ or $\div\disA$, or combinations thereof. One can find a differential that is capable of extracting local tensor modes. Such an extractor is precisely what was found in \cite{Clarkson:2011td} and takes the form:
 \ber\hat{\mathscr{T}}(Y_{ab})\equiv [-\sdel^2+2\cavK]\curlA (Y_{ab})\label{tensor-ex} \ear \footnote{$[-\sdel^2+2\cavK]$ can also be expressed as $\:13\left[\sdel^{2}+\curlA^{2}-2\cavK\right]$ \cite{Clarkson:2011td}} we note that $\curlA$ commutes with the operator in square brackets. Applying the extractor to our hypothetical wave equation, it is clear that 
\begin{eqnarray}\label{TINA}\hat{\mathscr{T}}(\mathcal{L}[Y_{ab}])=\mathcal{L}(\hat{\mathscr{T}}[Y_{ab}]) =\hat{\mathscr{T}}(\mathcal{W}_{ab}),\end{eqnarray} since $\hat{\mathscr{T}}$ commutes with $\mathcal{L}$. It is relatively straightforward to verify that $\hat{\mathscr{T}}(Y_{ab})$ and $\hat{\mathscr{T}}(\mathcal{W}_{ab})$ are transverse showing that the extractors yields pure tensors as required. To convert the extraction into Fourier space, one needs to define tensor harmonics $\sdel^2\Qt_{ab}=-k^2\Qt_{ab}$, where we have two parities of orthogonal harmonics, $(k^2+3\cavK)^{1/2}\Qt_{ab}=\curlA\Qtb_{ab}\Leftrightarrow (k^2+3\cavK)^{1/2}\Qtb_{ab}=\curlA\Qt_{ab}$. A Fourier composed form of (\ref{TINA}) takes the form \ber \left(k^{2}+3\cavK\right)^{1/2}\left(k^{2}+2\cavK\right)Y^{(k)}=\mathcal{W}^{(k)}, \ear for a unique wave number. We will elaborate this equality when we discuss the specific case later in the article. Let's now lay down the foundation for the specific case that we are interested in.
 \section{\label{Key} Key second order variables and equations } 
Order mixing and the mixing of perturbations types mentioned in section (\ref{HO}) present significant challenges to the analysis of second order perturbations. The utilities discussed above can help resolve some of these challenges but one has to be very clear about the starting point if any progress is to be made. In this regard, we begin with a narrowed down case, but one that will help illustrate how to deal with these challenges. It is entirely plausible and sensible to expect that one type of perturbations at say, first order, should be linked to the same type of perturbation at say second order. This is in deed true, but it also possible for first order scalar perturbations to be linked to second order vector or tensor perturbations i.e. one type of perturbations acting as a source for a different type of perturbations at a higher order of perturbations  (see the discussion of the {\it bottom-up} approach). In the {\it top-down} approach this says that second order tensor type may be desegregated into portions that link to a completely different type of perturbations at the first order. 

We look to 1+3 splitting and the extractor to define our basic variable. First note that our extractor is made up of pre-factor $ (-\sdel^2+2\cavK)$ multiplied by $\bar{\curl}(\equiv\mathscr{L} \curl$, where $\mathscr{L}$ is the scale-factor). If the pre-factor is labeled $\alpha$, then the extractor takes the simplified form $\alpha\mathscr{L} \curl$. It is clear that the $\alpha\mathscr{L}$ takes on a numerical value in Fourier space if the scale factor, the wave number and the curvature are specified. With this in mind, we can define a simplified second order gauge invariant quantity. We note that the vanishing of $\curl$ is at the core of the definition of the pure scalar part in the 1+3 splitting (see for example Eq. (\ref{SplitT}) and the accompanying explanation). We apply the extractor to first order shear tensor $\hat{\mathscr{T}}[\sigma_{ab}]$. This will yield a pure first order tensor because the extractor does not change the order of the quantity but rather extracts the pure tensor part of it. But now assume that only scalar perturbations are excited at first order then $\hat{\mathscr{T}}[\sigma_{ab}]=0$ at this order, but $\hat{\mathscr{T}}[\sigma_{ab}]\neq0$ at second order. This, using Stewart-Walker lemma, is a gauge-invariant quantity at upto second order. We denote this new quantity by $\Sigma_{ab}=\hat{\mathscr{T}}[\sigma_{ab}]$. Since we do not have vectors and tensors at first order ( remember we only excited scalar perturbations), it is sufficient to use $\curlA$ as a tensor extractor rather than the full $\hat{\mathscr{T}}$. This is because the pre-factor $(-\sdel^2+2\cavK)$, which appears before $\curlA$ in Eq. (\ref{tensor-ex}), is rank preserving while the whole extractor commutes with the time derivative and may be treated as a unit. For our purposes, it will be sufficient to use just the $\curl$ as an extractor as we explain above. In general though, one would need to apply the full extractor to obtain the tensor part. Since the general equation of state adds another level of complexity, we will restrict our discussion to two simple cases: that of 'dust' (p=0) (this has been considered elsewhere \cite{Clarkson:2003af}) and the barotropic equation state (p=w$\mu$). To the best of our knowledge, the latter development is new and fills a gap thereby allowing a complete formalism for perfect fluids. The case for imperfect fluid has not been developed and is reserved for future work. 

\section{ \label{dusty}The case of dust equation of state}
As discussed above, we consider second order pure tensor perturbations. We first note that  for 'dust' EOS w=0 ($\Rightarrow p=0$ and from Eq. (\ref{presCon}) that $A^{a}$=0). Now consider the special case where only linear order perturbations are excited at first order. Following our discussion in the previous section, $\curl\sigma_{ab}$ characterizes second order perturbations and $\alpha\mathscr{L}\curl\sigma_{ab}$ ( where again $\alpha$ is the pre-factor mentioned above and $\mathscr{L}$ the scale factor) represents pure second order tensor. Using our idea of {\it filters} we can group variables into background quantities ($\mu, \Theta$ and $p=0$), first order quantities (scalar parts of $X_{a},\mathcal{Z}_{a}, \sigma_{ab}$ and $E_{ab},$) and second-order quantities ($\curl\sigma_{ab}$). Let us  focus on the shear tensor, from constrain (Eq. \ref{H1}) it is clear that $\curl[{\sigma}_{ab}]\equiv H_{ab}$. This means that if we did not consider the restriction that only scalars are present at first-order, then $H_{ab}$ would be a first-order quantity that is equivalent to the part of the shear tensor that is devoid of scalars. But in our restricted case, this part vanishes at first order, since the curl sets the scalar part of the shear tensor to zero. It is obvious that  $\curl[\sigma_{ab}]$ has both vector and tensor parts, but we can obtain a pure tensor part by applying the tensor extractor to the wave equation involving this quantity. Let us consider the SOGI term,\be \label{chp6:2}\Sigma_{ab}=\curl\sigma_{ab}.\ee From Eq.(\ref{Alldot}), it can be shown that the time derivative of $\Sigma_{ab} (\equiv H_{ab})$ takes the form \be \label{chp6:3}\dot{\Sigma}_{ab}=-\Theta\Sigma_{ab}-\curl(E)_{ab}, \ee for the case where {\it second-order-filters} are applied coupled with the restrictions adopted in this is section, namely scalar perturbations at first-order. Now taking the time derivative of Eq. (\ref{chp6:3}) and using the commutation
relation Eq. (\ref{app:10}) gives
\ber\label{shearddot}\ddot{\Sigma}_{ab}+\curl\curl\Sigma_{ab}+\sfrac{7}{3}\Theta\dot{\Sigma}_{ab}+(\Theta^{2}-\mu)\Sigma_{ab}&=&\mathcal{F}_{ab},
\ear 
where \[\mathcal{F}_{ab}\equiv\epsilon_{cd\langle
a}[\sfrac{3}{2}{E_{b\rangle}}^{d}\mathsf{\mathsf{div}}(\sigma^{c})+\sfrac{3}{2}{\sigma_{b\rangle}}^{d}~D^{e}{E_{e}}^{c}+\sigma^{ec}~D_{|e|}{E_{b\rangle}}^{d}]-3\curl(\sigma_{c\langle a}{E_{b\rangle}}^{c}),\] where $D_{|e|}$ implies unprojected index. We know that the left hand side is gauge invariant by virtual of the definition of $\Sigma_{ab}.$ The relevant question to ask is if the source term is equally gauge invariant. In order to ascertain this, we have to demonstrate that $\mathcal{F}_{ab}=0$ when $\Sigma_{ab}=0$ (both in the background and at first-order {\it filtering}). There are several ways to achieve this but we only point out one in this article. We use the $\dot{E}_{ab}$ part of Eq.(\ref{Alldot}) to express the last term of $\mathcal{F}_{ab}$ as follows \begin{eqnarray}
\label{sec:18}
\curl\dot{E}_{ab}+\Theta\curl E_{ab}+ \sfrac{3}{2}\epsilon_{cd\langle
a}{E_{b\rangle}}^{d}\mathsf{div}(\sigma^{c})-\curl\curl\Sigma_{ab}+\sfrac{1}{2}\mu\Sigma_{ab}+\sfrac{3}{2}\epsilon_{cd\langle
a}{\sigma_{b\rangle}}^{d}\mathsf{\mathsf{div}}(E^{c})
\end{eqnarray} This new expression can now be substituted back in the source $\mathcal{F}_{ab}$ to give 
\begin{eqnarray}
\label{sec:19}\mathcal{F}_{ab}&=&-\curl\dot{E}_{ab}-\Theta\curl
E_{ab}+
\curl\curl\Sigma_{ab}-\sfrac{1}{2}\mu\Sigma_{ab}+\epsilon_{cd\langle
a}\sigma^{ec}~D_{|e|}{E_{b\rangle}}^{d},
\end{eqnarray} and on applying commutation relation Eq. (\ref{app:10}), one finds
\begin{eqnarray}
\mathcal{F}_{ab}&=&-(\curl E_{ab})^{.}-\sfrac{4}{3}\Theta\curl E_{ab}+\curl
\curl\Sigma_{ab}-\sfrac{1}{2}\mu\Sigma_{ab}.
\end{eqnarray} We know that both $\Sigma_{ab}$ and $\mathcal{F}_{ab}$ vanish in the background given our discussions in the preceding sections. Now note, from Eq. (\ref{chp6:3}), that $\Sigma_{ab}=0\Rightarrow \curl
E_{ab}=0$ and so $\mathcal{F}_{ab}$ vanishes at first-order as required, with the implication that $\mathcal{F}_{ab}$ is also gauge-invariant. What remains now is to solve the wave equation for $\Sigma_{ab}$ before applying the pure tensor extractor. First note that we need to determine how $\mathcal{F}_{ab}$ evolves, which is what is done in the next section.

\subsection{\label{sec:level1}Propagating the source term $\mathcal{F}_{ab}$} It is useful to express the terms in $\mathcal{F}_{ab}$ in some compact formulation that make it easier to handle. In this regard, we define the following new compact objects~\cite{Clarkson:2003af}.  \ber\psi_{1ab}&=&\epsilon_{cd\langle
a}\mathsf{div}(\sigma^{c}){\sigma_{b\rangle}}^{d},~~~~\xi_{1ab}=\epsilon_{cd\langle
a}\sigma^{ec}~D_{|e|}{\sigma_{b\rangle}}^{d},\nn\\
\psi_{2ab}&=&\epsilon_{cd\langle
a}\mathsf{div}(\sigma^{c}){\dot{\sigma}_{b\rangle}}^{d},~~~~~\xi_{2ab}=\epsilon_{cd\langle
a}\dot{\sigma}^{ec}~D_{|e|}{\sigma_{b\rangle}}^{d},\nn\\
\psi_{3ab}&=&\epsilon_{cd\langle
a}\mathsf{div}(\dot{\sigma}^{c}){\sigma_{b\rangle}}^{d},~~~~\xi_{3ab}=\epsilon_{cd\langle
a}\sigma^{ec}~D_{|e|}{\dot{\sigma}_{b\rangle}}^{d},\nn\\
\psi_{4ab}&=&\epsilon_{cd\langle
a}\mathsf{div}(\dot{\sigma}^{c}){\dot{\sigma}_{b\rangle}}^{d},~~~~~\xi_{4ab}=\epsilon_{cd\langle
a}\dot{\sigma}^{ec}~D_{|e|}{\dot{\sigma}_{b\rangle}}^{d}\label{NewVar1}
\ear It follows that $\mathcal{F}_{ab}$ takes the simple but compact form, \ber\label{here}\mathcal{F}_{ab}&=&-\sfrac{9}{2}\psi_{2ab}-4\Theta\psi_{1ab}-\sfrac{3}{2}\psi_{3ab}-4\xi_{3ab}-\sfrac{8}{3}\Theta\xi_{1ab}.
\ear  Taking the time derivative of each variable given in the array (\ref{NewVar1}), making use of the commutation relation Eq.(\ref{app:10}) and subjecting the resulting propagation equation to the second order filter yields the following closed set of first order differential equations,
\begin{widetext}
\ber
\label{psipro}
\left[
\begin{array}{c}
{\psi}_{1ab}  \\
{\psi}_{2ab} \\
{\psi}_{1ab}
\end{array}
\right]^{.}=
\left[
\begin{array}{cccc}
 -\frac{1}{3}\Theta & 1  & 1 & 0  \\
-(\frac{4}{9}\Theta^2-\frac{5}{6}\mu)  &  -2\Theta & 0 &1   \\
-(\frac{4}{9}\Theta^2-\frac{5}{6}\mu)  &  0 &-2\Theta &1   \\
 0 &  -(\frac{4}{9}\Theta^2-\frac{5}{6}\mu) & -(\frac{4}{9}\Theta^2-\frac{5}{6}\mu)&-\frac{11}{3}\Theta  
\end{array}
\right]\left[
\begin{array}{c}
{\psi}_{1ab}  \\
{\psi}_{2ab} \\
{\psi}_{1ab}
\end{array}
\right]
,\ear\end{widetext} 
It is clear that this set takes the form $\dot{\psi}_{ab}=\Psi{\psi}_{ab}, $ and can be solved relatively easily. Note that the propagation of $\psi$ is structurally similar to that of $\xi$ and so the $\xi$'s will yield as system of the form $\dot{\xi}_{ab}=\Xi \xi_{ab}$, where $\Xi=\Psi$. It is easy to obtain the solution for the differential system given in Eq. (\ref{psipro}). Note that one only needs to specify the values of quantities $\Theta$ and $\mu$ in the background. For a flat FLRW background the scale-factor $\mathscr{L}=\mathscr{L}_{0}t^{2/3}$ ($\Rightarrow\Theta=2/3t$ and $\mu=4/9 t^{2}$) where $t$ is the proper time. Substituting these into the system, integrating, isolating the relevant quantities and plugging them into $\mathcal{F}_{ab}$, Eq. (\ref{here}) yields\cite{Clarkson:2003af}, \be
 \label{SolF}\mathcal{F}_{ab}=\alpha_{ab}t^{-4}+\beta_{ab}t^{-\sfrac{7}{3}}+\gamma_{ab}t^{-\sfrac{17}{3}},\ee
 where ($\alpha_{ab}$, $\beta_{ab}$ and $\gamma_{ab}$) are the coefficients determined by the initial conditions. We present in appendix section (\ref{Alter}) an alternative approach of handling the source term. What remains now is the solution of the non-homogeneous wave equation for $\Sigma_{ab}.$

\subsection{Solutions for the dust sub-case}From Eqs. (\ref{SolF}, \ref{shearddot} ) and the commutation relation Eq. (\ref{app:11}), it follows that \ber\label{SHEAR}\ddot{\Sigma}_{ab}-~D^{2}\Sigma_{ab}+\sfrac{7}{3}\Theta\dot{\Sigma}_{ab}+(\Theta^{2}-\mu)\Sigma_{ab}&=&\alpha_{ab}t^{-4}+\beta_{ab}t^{-\sfrac{7}{3}}+\gamma_{ab}t^{-\sfrac{17}{3}}.\ear As in the previous section, both $\mu$ and $\Theta$ can be expressed in terms of the proper time $t$. We require a 1+3 harmonics decompositions of this equation, for us to be in a position to analyze it. The details of these decompositions can be found in ~\cite{RevModPhys.39.862,EHvE98,Bruni:1992dg, Challinor:1999xz}. Since $\Sigma_{ab}$ and $\mathcal{F}_{ab}$ are functions of time and space, they should ideally be written as $\Sigma_{ab}(t,x)$ and $\mathcal{F}_{ab}(t,x)$. The harmonic decomposition of these variables into temporal and spatial dependent parts, for the two parities ( electric and magnetic), take the form; \be\Sigma_{ab}(t,x)=\mathscr{L}^{-2}\sum_{\kappa}\kappa^{2}\left[\Sigma^{(\kappa)}(t)Q_{ab}(\kappa,x)+\bar{\Sigma}^{(\kappa)}(t)\bar{Q}_{ab}(\kappa,x)\right],\ee and \be\mathcal{F}_{ab}(t,x)=\mathscr{L}^{-2}\sum_{\kappa}\kappa^{2}\left[\mathcal{F}^{(\kappa)}(t)Q{_{ab}(\kappa,x)}+\bar{\mathcal{F}}^{(\kappa)}(t)\bar{Q}{_{ab}(\kappa,x)}\right].\ee It follows that the decomposition of (\ref{SHEAR}) reduces to, \ber\label{SHEAR1}\ddot{\Sigma}^{(\kappa)}+\sfrac{7}{3}\Theta\dot{\Sigma}^{(\kappa)}+(\sfrac{\kappa^{2}}{\mathscr{L}^2}+\Theta^{2}-\mu)\Sigma^{(\kappa)}&=&\mathcal{F}^{(\kappa)}, \ear for one parity, the other parity can similarly be written down. There is hidden parity switch due to the effect of $\curl$ that is subsumed in this equation. In particular \[\curl Q_{ab}=\sfrac{\kappa}{\mathscr{L}}\bar{Q}_{ab},\] for a flat model. It is clear that the double $\curl$ in Eq.(\ref{shearddot}) auto-corrects this parity switch. It should be noted that the two parities would generally couple. Our decoupled case takes the form 
\ber\label{SHEAR2}\ddot{\Sigma}^{(\kappa)}+\sfrac{28}{9 t^{2}}\dot{\Sigma}^{(\kappa)}+(\sfrac{\kappa^{2}}{{{\mathscr{L}_{0}}}^{2}t^{4/9}}+\sfrac{8}{27 t^{2}})\Sigma^{(\kappa)}&=&\mathcal{F}^{(\kappa)}, \ear when the background variables are expressed in terms of the proper time.
The challenge now is to pick the matching wave number for both the homogeneous part of the wave equation and the corresponding source. In particular, we chose identical $\kappa$. It is straight forward to show that the solutions to the homogeneous part of the wave equation (\ref{SHEAR1}) are, \be
 \Sigma_{1}^{(\kappa)} = C_{1}(\kappa)t^{-\sfrac{11}{6}}{\bf J}\left(\sfrac{5}{6},\sfrac{\kappa }{\mathscr{L}_{0}}t\right),~~~ \Sigma_{2}^{(k)}=C_{2}(\kappa)
  t^{-\sfrac{11}{6}}{\bf Y}\left(\sfrac{5}{6},\sfrac{\kappa }{\mathscr{L}_{0}}t\right),
 \ee where $C_{1}(\kappa)$ and $C_{2}(\kappa)$ are the constants of integration, while {\bf J}$(\equiv Bessel J)$ and {\bf Y}$(\equiv Bessel Y)$ are the Bessel functions of the first and the second kinds respectively. The general solutions can then be found using Green's method as follows,
 \ber\Sigma^{(\kappa)} = C_{1}(\kappa)
t^{-\sfrac{11}{6}}{\bf J}\left(\sfrac{5}{6},\sfrac{\kappa }{a}t\right)+C_{2}(\kappa)
  t^{-\sfrac{11}{6}}{\bf Y}\left(\sfrac{5}{6},\sfrac{\kappa }{a}t\right)+\sfrac{\pi}{2}\Sigma_{1}^{(\kappa)}\int{\Sigma_{2}^{(\kappa)}\mathcal{F}^{(\kappa)}}dt+\sfrac{\pi}{2}\Sigma_{2}^{(\kappa)}\int{\Sigma_{1}^{(\kappa)}\mathcal{F}^{(\kappa)}}dt.\nn\\ \ear For completeness, we also present numerical solutions to the wave equation in {figure (\ref{figdust})} for selected initial conditions. 
\begin{figure}[htbp]
\begin{center}
   \includegraphics[width=2.5in]{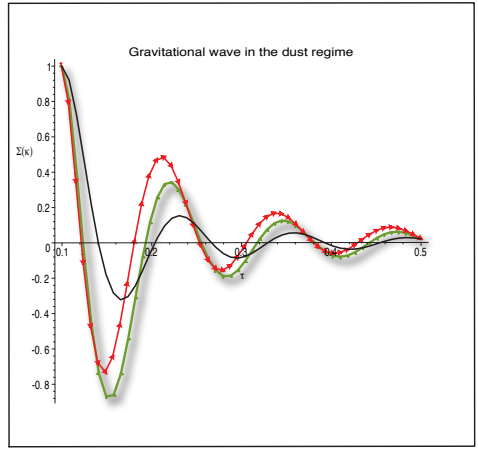}
\caption{\label{figdust}\small{{\it These curves represent numerical solutions to the gravitational wave equation (\ref{SHEAR1}) given different scaling of coefficients of the source term. The initial conditions are $\Sigma(\kappa)={\Sigma^{(\kappa)}}/\Sigma^{(\kappa)}_{0}=1, {\Sigma(\kappa)}'=0$, $\sfrac{\kappa}{\mathscr{L}_{0}}=25$. The red curve (with arrow head) represents the numerical solution for the homogeneous wave. The green curve (with dots) is the solution with $\alpha=0,\beta=.08,\gamma=-0.03$. The black curve is the solution curve for GW whose source has the coefficients; $\alpha=0.1,\beta=0, \gamma=-0.02$.} }}
\end{center}
\end{figure}The results in this section recover those found in ~\cite{Clarkson:2003af}. We now turn to $w\neq 0$ sub-case, where we obtain new results. 

\section{\label{Bar}Barotropic perfect fluid}
 \par The general non-vanishing pressure( $p\neq0$) introduces a significant level of complexity. Progress can however be made for the barotropic case ($p=\mathsf{w}\mu$). The starting point is similar to that presented in the previous section where 'dust' ($p=0$) was considered. Here too we begin with proviso that only scalar perturbations are excited at first-order and similarly, $\Sigma_{ab}$ is second order and gauge invariant. In this case the background quantities are: $\mu, \Theta, p=\mu(1+\mathsf{w})$, first-order quantities are the scalar parts of $X_{a},\mathcal{Z}_{a}, \sigma_{ab},E_{ab}$, $A_{a}$ and the second-order quantity is $\curl\sigma_{ab}$. The presence of pressure introduces an extra term in the source to the propagation equation for $\Sigma_{ab}$. In particular,
\be\label{Nshear:1}\dot{\Sigma}_{ab}+\Theta\Sigma_{ab}+\curl(E)_{ab}=-2\epsilon_{cd\langle
a}A^{c}{E_{b\rangle}}^{d},
\ee where we have used $A^{c}\equiv\dot{u}^{c}\neq0$\footnote{This should not be confused with the $\mathcal{A}_{ab}$ used in (\ref{Avar})}. Using the momentum conservation equation (\ref{presCon}) and the constraint involving the divergence of the electric part of the Weyl tensor from (\ref{const1}), we can now express $A^{c}$ in the following form,
\be\label{meq} A^{c}=-\frac{~D_{a}p}{(\mu+p)}=-\frac{\mathsf{w}~D_{a}\mu}{\mu(1+\mathsf{w})}=-\frac{3\mathsf{w}~D_{e}E^{ec}}{\mu(1+\mathsf{w})},\ee  
where $\mathsf{w}$ is constant is a true constant and takes on numerical values. Taking the time derivative of (\ref{Nshear:1}), using the commutation relation Eq. (\ref{app:10}) and applying the {\it second-order-filter} gives 
\ber\label{Nshearddot}\ddot{\Sigma}_{ab}+\curl\curl\Sigma_{ab}+\sfrac{7}{3}\Theta\dot{\Sigma}_{ab}+[\Theta^{2}-\mu(1+2\mathsf{w})]\Sigma_{ab}
&=& \mathcal{F}_{ab},\ear where $\mathcal{F}_{ab}$ is given by \ber&&\epsilon_{cd\langle a}[\sfrac{9}{2}{E_{b\rangle}}^{d}\mathsf{div}(\sigma^{c})+\sfrac{3}{2}(1+\mathsf{w}){\sigma_{b\rangle}}^{d}~\mathsf{div}(E)^{c}+ 4\sigma^{ec}~D_{|e|}{E_{b\rangle}}^{d}]\nn\\&&~~~~~~~~~~~~~~~~~~~~~~~~~+\sfrac{15\mathsf{w}}{\mu(1+\mathsf{w})}\epsilon_{cd\langle
a}{\dot{E}_{b\rangle}}\mathsf{div}(E^{c})+\sfrac{(13+6\mathsf{w})\mathsf{w}}{\mu(1+\mathsf{w})}\Theta\epsilon_{cd\langle
a}{{E}_{b\rangle}}\mathsf{div}(E^{c})\nn\\\ear It is necessary to express the source $\mathcal{F}_{ab}$ in terms of one of either $\sigma$ or $E$. This will allow us to construct variables analogous to those in Eq. (\ref{psipro}). We elect to eliminate the shear tensor. Since $\sigma$ appears in $\mathcal{F}_{ab}$ as part of a product (e.g $\sigma E$), we use Eq. (\ref{prop1}) to express $\sigma$ in terms of $E$. It follows that ${\sigma_{b}}^{d}$ and $\mathsf{div}(\sigma)^{c}$ can be expressed as \ber {\sigma_{b}}^{d}&=&-\frac{2({\dot{E}_{b}}^{d}+\Theta {E_{b}}^{d})}{\mu(1+\mathsf{w})},\\ \mathsf{div}\sigma^{c}&=&-\frac{2(\mathsf{div}(\dot{E})^{c}+\Theta \mathsf{div}E^{c})}{\mu(1+\mathsf{w})}.\ear Although these expressions do not apply when $\mathsf{w}=-1$, they suffice for the case we are interested in. For now, our substitutions lead to a source that is made up of products that involve only the electric part of the Weyl tensor. These are $\epsilon_{cd\langle a}{{E}_{b\rangle}}^{d}\mathsf{div}E^{c}$ and $
\epsilon_{cd\langle a}E^{ec}~D_{|e|}{E_{b\rangle}}^{d}$ and their temporal derivatives. It is easily to demonstrate that the propagation of these two terms give rise to two separate, but structurally similar systems. The solutions are also structurally identical and for this reason, will only develop and analyze the part involving $\epsilon_{cd\langle a}{{E}_{b\rangle}}^{d}\mathsf{div}E^{c}$. The source $\mathcal{F}_{ab}$ equals \be\label{sogisor}
\left(\frac{-20+12\mathsf{w}}{\mu(1+\mathsf{w})}\epsilon_{cd\langle
  a}{\dot{E}_{b\rangle}}^{d})\nn\\-\frac{20-10\mathsf{w}-6\mathsf{w}^2}{\mu(1+\mathsf{w})}\Theta \epsilon_{cd\langle
  a}{E_{b\rangle}}^{d}\right)\mathsf{div}(E^{c}).\ee As in the previous section, can now introduce compact objects that are products of first-order quantities. In particular, we define and make use of the following new  second-order variables,
  \ber\label{ps1} \psi_{1ab} &=&\epsilon_{cd\langle a}{E_{b\rangle}}^{d}\mathsf{div}(E)^{c},~~~~\label{xi1} \xi_{1ab} =\epsilon_{cd\langle a}E^{ec}~D_{|e|}{E_{b\rangle}}^{d}\\
  \label{ps2}\psi_{2ab} &=&\epsilon_{cd\langle a}{\dot{E}_{b\rangle}}^{d}\mathsf{div}(E)^{c},~~~~~\xi_{2ab} =\epsilon_{cd\langle a}\dot{E}^{ec}~D_{|e|}{E_{b\rangle}}^{d},\\
 \label{ps3}\psi_{3ab} &=&\epsilon_{cd\langle a}{\dot{E}_{b\rangle}}^{d}\mathsf{div}(\dot{E})^{c},~~~~~\xi_{3ab} =\epsilon_{cd\langle a}\dot{E}^{ec}~D_{|e|}{\dot{E}_{b\rangle}}^{d},\\
 \label{ps4} \psi_{4ab} &=&\mu\epsilon_{cd\langle a}{L_{b\rangle}}^{d}\mathsf{div}(E)^{c},~~\xi_{4ab} =\epsilon_{cd\langle a}L^{ec}~D_{|e|}{E_{b\rangle}}^{d}\\
  \label{ps5}\psi_{5ab} &=&\mu\epsilon_{cd\langle a}{L_{b\rangle}}^{d}\mathsf{div}(\dot{E})^{c}~~~\xi_{5ab} =\epsilon_{cd\langle a}E^{ec}~D_{|e|}{\dot{E}_{b\rangle}}^{d},\\
  \label{ps6}\psi_{6ab} &=&\mu\epsilon_{cd\langle a}{L_{b\rangle}}^{d}\mathsf{div}(L)^{c}~~~~\xi_{6ab} =\epsilon_{cd\langle a}L^{ec}~D_{|e|}{L_{b\rangle}}^{d},\ear where $L_{ab}=~D_{\langle a}A_{b\rangle}.$ Whereas the first two terms appear in the source term $\mathcal{F}_{ab},$ the remaining arise in the propagation of the different terms that make up the source. 
 \begin{widetext} 
 
 \subsection{Analyzing the source $\mathcal{F}_{ab}$ for $\mathsf{w}\neq0$ and $\mathsf{w}\neq-1$ }
We will use a method similar to that employed in the analysis of the source term for the {\it dust} case. The propagation of the variables (\ref{ps1}-\ref{ps6}), using the relevant relations in Eq. (\ref{prop1}) and applying a {\it second-order-filter} yields a system of the form \[\dot{\psi}_{n}=\Psi\psi_{n}+\phi_{n},\] 
 where $n=1..6$ and $\Psi$ is the coupling matrix given by
\[\Psi= \left[
  \begin{array}{cccccc}
    -\sfrac{1}{3}\Theta & 2 & 0 & 0 & 0 & 0 \\
    -(1+\sfrac{1}{3}\mathsf{w})\Theta^2 & -(3+\mathsf{w})\Theta & 1 & -\sfrac{1}{2}(1+\mathsf{w}) & 0 & 0\\
    0 & -2(1+\sfrac{1}{3}\mathsf{w})\Theta^2 & -(\sfrac{17}{3}+2\mathsf{w})\Theta & 0 & -(1+\mathsf{w}) & 0 \\
    0 & 0 & 0 & -2\Theta & 1 & 0 \\
    0 & 0 & 0 & -(1+\sfrac{1}{3}\mathsf{w})\Theta^2 & -(\sfrac{14}{3}+\mathsf{w})\Theta & -\sfrac{1}{2}\mu(1+\mathsf{w}) \\
    0 & 0 & 0 & 0 & 0 & -(\sfrac{8}{3}-\mathsf{w})\Theta
  \end{array}
\right],\] while the residual components are give by the column matrix
\[
\phi_{n}=\left[
\begin{array}{c}
 0 \\
 0 \\
0\\   
\mathsf{w}\mu \epsilon_{cd\langle a}~D^{2}{\sigma_{b\rangle}}^{d}\mathsf{div}(E)^{c} \\
\mathsf{w}\mu \epsilon_{cd\langle a}~D^{2}{\sigma_{b\rangle}}^{d}\mathsf{div}(\dot{E})^{c} \\
-\mathsf{w}\Theta\psi_{6ab}+\mathsf{w}\mu \epsilon_{cd\langle a}{L_{b\rangle}}^{d}~D^{2}\mathsf{div}(\sigma)^{c}+2\mathsf{w}\mu \epsilon_{cd\langle a}{~D^{2}\sigma_{b\rangle}}^{d}\mathsf{div}(L)^{c}
\end{array}
\right]
\] We now apply the standard 1+3 harmonic decomposition to system of equations. Again we only present the case for one parity. In order to do this, we introduce a new notation for the harmonic decomposition of our new variables. For example,
\ber\psi_{1ab}=\sum_{\kappa=\kappa'}E_{1}(\kappa)\epsilon_{cd\langle a}{Q_{b\rangle}}^{(k)d}\mathsf{div}[E_{1}(\kappa') Q^{(\kappa')c}]=\sum_{\kappa=\kappa'}\psi_{1}(\kappa,\kappa')\epsilon_{cd\langle a}{Q_{b\rangle}}^{d}(k)\mathsf{div}Q^{c}(\kappa').\ear All the terms in the above system can be decomposed similarly. First notice that \[\mathsf{w}\mu\epsilon_{cd\langle a}{~D^{2}\sigma_{b\rangle}}^{d}\mathsf{div}(E)^{c}=\frac{2\mathsf{w}}{(1+\mathsf{w})}\epsilon_{cd\langle a}{~D^{2}\left({\dot{E}_{b}}^{d}+\Theta {E_{b}}^{d}\right)\mathsf{div}(E)^{c}}\] and hence,\ber \label{lapp1} \frac{2\mathsf{w}}{(1+\mathsf{w})}\epsilon_{cd\langle a}{~D^{2}\left({\dot{E}_{b}}^{d}+\Theta {E_{b}}^{d}\right)\mathsf{div}(E)^{c}}&=&\frac{2\mathsf{w}k^{2}}{(1+\mathsf{w})\mathscr{L}^{2}}[\psi_{2}(\kappa,\kappa')+\psi_{1}(\kappa,\kappa')]\epsilon_{cd\langle a}{Q_{b\rangle}}^{(k)d}\mathsf{div}Q^{(\kappa')c},\nn\\\mathsf{w}\mu\epsilon_{cd\langle a}{~D^{2}\sigma_{b\rangle}}^{d}\mathsf{div}(\dot{E})^{c}&=&\frac{2\mathsf{w}k^{2}}{(1+\mathsf{w})\mathscr{L}^{2}}[\psi_{2}(\kappa,\kappa')+\psi_{1}(\kappa,\kappa')]\epsilon_{cd\langle a}{Q_{b\rangle}}^{d}(\kappa)\mathsf{div} Q^{c}(\kappa')\nn\\\mathsf{w}\mu\epsilon_{cd\langle a}{~D^{2}\sigma_{b\rangle}}^{d}\mathsf{div}(L)^{c}+2\mathsf{w}\mu\epsilon_{cd\langle a}{L_{b\rangle}}^{d}~D^{2}\mathsf{div}(\sigma)^{c}&=&\frac{4\mathsf{w}k^{2}}{(1+\mathsf{w})\mathscr{L}^{2}}[\psi(\kappa,\kappa')_{4}+2\psi(\kappa,\kappa')_{5}]\epsilon_{cd\langle a}{Q_{b\rangle}}^{d}(\kappa)\mathsf{div}Q^{c}(\kappa').\nn\\\ear One could attempt to solve the system in Fourier space and then use this to solve the wave equation in Fourier space, but physically motivated considerations reduce the amount of algebra required to make progress and that is what we need. In this case, we only need the long wave length limit of the second order quantities represented by these compact objects. The motivation comes the inflationary and the curvaton scenarios in the early universe where the wavelength of cosmological perturbations responsible for seeding the present cosmic structures is need be much larger than the horizon scales. We similarly argue that the long wavelength limit, where these terms effectively vanish, will be responsible for seeding an enduring second order wave equation. The net effect is that the system reduces to the homogeneous case $\dot{\psi}_{n}=\Psi\psi_{n}$. Note that setting both $\theta$ and $\mu $ to their background values i.e. $\Theta=\sfrac{2}{3(1+\mathsf{w})t}$ allows the homogeneous system to be easily analyzed. 
This will yield values for each of the terms given in (\ref{ps1}-\ref{ps6}). Substituting the solutions obtained for $\psi_{1ab}$ and $\psi_{2ab}$ into (\ref{sogisor}) shows that $\mathcal{F}_{ab}$ is equal to
\ber
\alpha_{1_{ab}}t^{-\sfrac{4}{(1+\mathsf{w})}}+\alpha_{2_{ab}}t^{\sfrac{2(-2+\mathsf{w})}{(1+\mathsf{w})}}+\alpha_{3_{ab}}t^{\sfrac{2(-2+3\mathsf{w})}{3(1+\mathsf{w})}}+\alpha_{4_{ab}}t^{\sfrac{(-7+9\mathsf{w})}{3(1+\mathsf{w})}}+\alpha_{5_{ab}}t^{\sfrac{(-7+15\mathsf{w})}{3(1+\mathsf{w})}} +\alpha_{6_{ab}}t^{-\sfrac{(17+3\mathsf{w})}{3(1+\mathsf{w})}}\nn\\\ear
\end{widetext}
 where
\ber\alpha_{1_{ab}}&=& \sfrac{5c_{1_{ab}}}{4}(-2+\mathsf{w})(-248+112\mathsf{w}+72\mathsf{w}^2),\\
\alpha_{2_{ab}}&=& \sfrac{5c_{2_{ab}}}{4}(-2+\mathsf{w})(-253+110\mathsf{w}+75\mathsf{w}^2),\ear \ber
\alpha_{3_{ab}}&=& \sfrac{5c_{3_{ab}}}{4}(-2+\mathsf{w})(-253+104\mathsf{w}+69\mathsf{w}^2),\\
\alpha_{4_{ab}}&=& \sfrac{5c_{4_{ab}}}{4}(-2+\mathsf{w})(-248+130\mathsf{w}+90\mathsf{w}^2),\ear \ber
\alpha_{5_{ab}}&=& \sfrac{5c_{5_{ab}}}{4}(-2+\mathsf{w})(-248+118\mathsf{w}+78\mathsf{w}^2),\\
\alpha_{6_{ab}}&=& \sfrac{5c_{6_{ab}}}{4}(-2+\mathsf{w})(-258+96\mathsf{w}+66\mathsf{w}^2),
\ear and where \(c_{n_{ab}}, n=1,2...6\) are constants of integration that are determined by the initial conditions. Now applying the standard 1+3 tensor harmonic decomposition to the second-order wave equation gives,
\ber
 \ddot{\Sigma}^{(\kappa)}+\sfrac{\kappa^{2}}{\mathcal{L}^{2}}\Sigma^{(\kappa)}+\sfrac{7}{3}\Theta\dot{\Sigma}^{(\kappa)}+\sfrac{2}{3}\Theta^{2}(1-\mathsf{w})\Sigma^{(\kappa)}&=& \mathcal{F}^{(k)}.\nonumber\\\ear for each parity. We first used commutation (\ref{app:11}) to convert the double $\curl$ into the Laplacian operator $-D^{2}$. The solutions to the homogeneous part are
\ber
 \Sigma_{1}^{(k)} =C_{1}(\kappa) t^{-\sfrac{11-3\mathsf{w}}{6(1+\mathsf{w})}}{\bf J}\left(\sfrac{5+3\mathsf{w}}{6(1+\mathsf{w})},\sfrac{\kappa }{\mathscr{L}_{0}}t\right),\\\Sigma_{2}^{(\kappa)}=C_{2} t^{-\sfrac{11-3\mathsf{w}}{6(1+\mathsf{w})}}\mathcal{\bf Y}\left(\sfrac{5+3{w}}{6(1+\mathsf{w})},\sfrac{\kappa }{\mathscr{L}_{0}}t\right),
 \ear  where again ${\bf J}$ and ${\bf Y}$ are the Bessel functions of the first and the kinds respectively. Here too, the general solutions can be found using Green's method i.e.
\begin{widetext} \ber\Sigma^{(k)} &=&
\Sigma_{1}^{(k)}+\Sigma_{2}^{(k)}+\Sigma_{1}^{(k)}\int{\sfrac{\mathscr{L}_{0}}{\varpi
k}\Sigma_{2}^{(k)}S^{(k)}}dt+\Sigma_{2}^{(k)}\int{\sfrac{\mathscr{L}_{0}}{\varpi
k}\Sigma_{1}^{(k)}S^{(k)}}dt
 \ear where \ber
\varpi&=&{\bf J}\left(\sfrac{(5+3w)}{6(1+\mathsf{w})},\sfrac{\kappa }{\mathscr{L}_{0}}t\right){\bf Y}\left(\sfrac{(11+9w)}{6(1+\mathsf{w})},\sfrac{\kappa }{\mathscr{L}_{0}}t\right)-{\bf Y}\left(\sfrac{(5+3w)}{6(1+\mathsf{w})},\sfrac{\kappa }{\mathscr{L}_{0}}t\right){\bf J}\left(\sfrac{(11+9w)}{6(1+\mathsf{w})},\sfrac{\kappa}{\mathscr{L}_{0}}t\right).\nn\\
\ear
\end{widetext} Figure (\ref{figGW}) gives the comparison of the numerical solutions for the second-order shear wave equation given dust $\mathsf{w}=0$ to that given radiation $\mathsf{w}=1/3$. The analytical and numerical solutions demonstrate that such a complex system in tractable and results interpreted with the context and restrictions used. But we have to ask, how relevant would such analysis be to the broader field of cosmology? To answer this one would have to look at aspects of cosmology where second order effect might play a role. We do this in the next section.
\begin{figure}[htbp]
\includegraphics[width=5in]{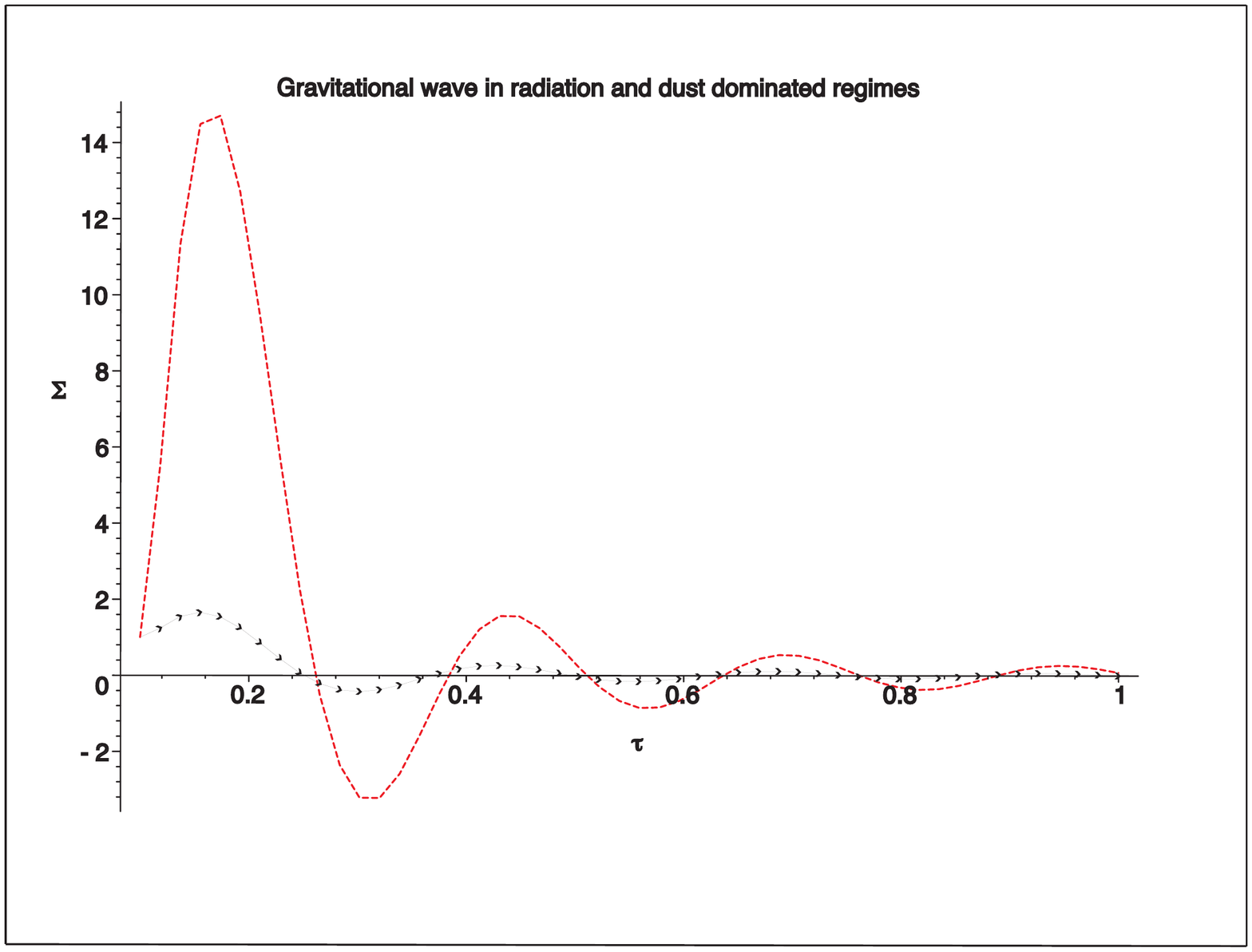}
\caption{ \small{{\it The red (dotted) line is that of second-order gravitational wave amplitude $\Sigma=\Sigma^{(\kappa)}/\Sigma^{(\kappa)}_{0}$ in the radiation regime, while the black line that in the matter regime. It is clear that the wave decays much faster in the radiation regime. The initial conditions are $\Sigma^{(\kappa)}_{0}=0.00001,~\Sigma^{(\kappa)'}=0,~k/a=25,~\alpha=\beta=\gamma=0.1$}}}
\label{figGW}

\end{figure} .
\section{\label{CMB}Second effects and CMB polarization}
It is known that any mechanism which produces temperature anisotropies such as perturbations will leave nonzero polarization of the CMB \cite{Bond1, Bond2, Zald, Plon, Critten, Harari, Koswosky, Mel}. The contribution of linear order perturbations to the polarization have been characterized \cite{Kam} and studied in great detail. 

As we pointed out in the abstract and in the introduction one would like to determine how a mathematical characterization of second effects such as the second-order tensors perturbations given in the previous section relates to effects on the CMB polarization. Since we have given the second order tensor perturbations in the 1+3 formalism, it would make sense to consider CMB polarization from the point of view of the 1+3 formalism. In the 1+3 covariant
approach the local angular distribution of the radiation is analyzed in terms of the projected symmetric and trace free (PSTF) tensor-valued multi-poles, which eliminates the need for harmonic functions. The resulting equations allow the analysis of the evolution of anisotropies and polarization for a general perturbation. We first briefly review the 1+3 decomposition techniques suited to the study of CMB anisotropies and polarization and then clearly point out how the above second order perturbations might show up in the polarization measurements. For a complete description of the formalism the reader is referred to \cite{Gebb}. For a comparative metric approach the reader is referred to \cite{BMMR}. We follow (Challinor's approach). Our description is based on the observation made by an observer co-moving the 4-velocity $u^{a}$. We first note that there exist degrees of freedom in the way $u^{a}$ can be chosen for a general cosmological model. To eliminate these degrees of freedoms we define $u^a$ to coincide with the time- like eigenvector of the matter stress-energy tensor or the 4-velocity of some particle species. This restriction on $u^a$ is necessary to ensure gauge-invariance of the 1+3 covariant perturbation theory. 

Now consider a photon having 4-momentum, ${P}^a$, ( not to be confused with the notation for pressure p) and energy, $E$ ( again not to be confused with the notation for the electric part of the Weyl tensor). Let the direction, relative to $u^{a}$, in which the photon is propagation be given by $e^{a}$. It is clear that 
 \[P^{a}=E(u^{a}+e^{a}).\] 
 Let $(e_{1})^{a}$ and $(e_{2})^{a}$ be two orthonormal ( a standard consideration in the 1+3 formalism) pair of polarization vectors that are orthogonal to both $u^{a}$ and $e^{a}$. One can define a projection tensor $\mathcal{H}_{ab}=h_{ab}+e_{a}e_{b}$ such that $(e_{1})_{a}=\mathcal{H}_{a}^{b}(e_{1})_{b}$. We set ($u^{a}, e^{a}, {e_{1}}^{a}, {e_{2}}^{a} $) to be right-handed orthonormal tetrad at the observation point. Using the polarization basis vector, the observer can decompose any vector field into Stokes parameters ($I, Q, U, V$), all functions of photon energy ($E$) and photon direction of propagation ($e^{a}$). Considering the way Stokes parameters transform under rotation of vectors $(e_{1})^{a}$ and $(e_{2})^{a}$, one can define the polarization tensor, $P_{ab} (E, e^{a})$ where the only non vanishing component is given by
\[ P_{ab}(e_{i})^{a}(e_{j})^{b}=\frac{1}{2} \left( \begin{array}{cc}
I+Q & U+V \\
U-V & I-Q \\
 \end{array} \right),\] and where $i$=1,2, and $j$=1,2. The full tensor in terms of $E$ and $e^{a}$ can be expressed in the form 
 \[P_{ab} (E, e^{a})=-\frac{1}{2}I \mathcal{H}_{ab}+\mathscr{P}_{ab}+\frac{1}{2}V {\epsilon_{abc}}e^{c},\] where $\mathscr{P}$is the linear polarization. $I$, $\mathscr{P}$ and $V$ are functions of $E$ and $e^{a}.$ The linear polarization tensor can be split into electric and magnetic parts i.e.
 \[\mathscr{P}_{ab}(E, e^{a})=\mathcal{E}_{ab}(E, e^{a})+\mathcal{M}_{ab}(E, e^{a})\]
For linear order perturbations, scalar perturbations contribute to the electric part of the polarization tensor, while tensor and vector perturbations contribute to both. The perturbation-type mixing and order mixing we encountered when looking at second order perturbations will invariably have an impact on the polarization tensor.  Take the magnetic part of the polarization tensor for example. One can express it in the form $\mathcal{M}_{ab}=\mathcal{M}^{(1)}_{ab}+\mathcal{M}^{(2)}_{ab}$, where the superscripts (1) and (2) denote the order of contributing tensor perturbation. The exact expression and the corresponding power spectra will be presented in the upcoming article \cite{bob2017}. It is important to note that the challenge is in devising techniques that could segregate the two contributions given the level of noise and the low level of polarization in CMB.  
\section{\label{Conc}Conclusion}
The interplay between theory and observations has spurred on interest in the the field of cosmology. We are witnessing a period where observations is driving this whole field of study, thanks to the abundance of data from various experiments. But with the abundance comes the need for precision, and with precision new questions. Perhaps the improving precision is the greatest feature of this data analysis in this field. In principle, this should allow us to determine the role played by nonlinear effects in cosmology, primordial or otherwise. One of the greatest hindrances to the study of nonlinear relativistic effects is mathematical, for this reason perturbation theory remains the most promising recourse. However, nonlinear perturbation theory is made even more complicated by gauge issues and complex equations. We have discussed a new second-order gauge invariant formalism suited for studying models with barotropic fluids. Our work extends   and provides alternative analytical tools to those presented in ~\cite{Clarkson:2003af}.
\par As observed in the ~\cite{Clarkson:2003af}, the main difficulty encountered in the 1+3 covariant approach at second-order is linked to the gauge-issue. Whereas equation (\ref{sogisor}) can be written by crossing off third-order terms, they are not in general integrable. This has to do with the fact that unlike the metric approach which solves for operators at first order, the covariant approach only solves for physical variables. In order to integrate the second-order equations therefore, derivative operators must operate only on variables which vanish at first-order and in the background. We have considered the the hypothetical case where only scalar perturbations are excited at linear. This has allowed the characterization of a second-order gauge invariant variables $\Sigma_{ab}, \psi_{(n)ab}, \xi_{(n)ab}$. $\Sigma_{ab}$ is equivalent to the tensor part of first-order $\sigma_{ab}$.  The $\psi_{(n)ab}$ and $\xi_{(n)ab}$, which make up the source for the $\Sigma_{ab}$ wave equations, are products of first-order scalars. These second-order gauge-Invariant  (SOGI) variables highlight the two different ways in which second-order gauge invariant variables may arise. From the numerical solutions to the $\Sigma_{ab}$ tensor wave equation, we find that the magnitude of the tensor is much greater in the radiation dominated universe than the dust dominated. Since tensor perturbations contribute to CMB polarization, we have provided a possible way of characterizing  such a contribution and will build on this in a future article \cite{bob2017}. 
\appendix
 
\section{\label{Alter}Alternative method for handling the source $\mathcal{F}_{ab}$ for the {\it dust case}}
 An alternative method for determining $\mathcal{F}_{ab}$ involves taking the time derivative of (\ref{here}), expressing the resulting equation in terms of the original variable and isolating a new source;
 \be \label{alta}\mathcal{\dot{F}}_{ab}+2\Theta\mathcal{F}_{ab}=\mathcal{A}_{ab},\ee where \ber\mathcal{A}_{ab}=\sfrac{11}{3}\Theta^{2}\psi_{1ab}+6\psi_{4ab}+4\Theta(\psi_{2ab}+\psi_{3ab}+\xi_{4ab})+\sfrac{8}{3}\Theta(\xi_{2ab}+\xi_{3ab}).\nn\\\ear Now taking the time derivative of (\ref{alta}), using the propagation equation each $\psi$s and $\xi$s as prescribed by the commutation relation (\ref{app:10}) and the {\it second order filter} yields:
\be\label{Avar}\dot{\mathcal{A}}_{ab}+\sfrac{8}{3}\Theta\mathcal{A}_{ab}=\mathcal{B}_{ab}, \ee where \ber\label{crap}\mathcal{B}_{ab}&=&-2\Theta(\psi_{4ab}+\frac{2}{3}\xi_{4ab})-\frac{10}{9}\Theta^{2}(3\psi_{3ab}+3\psi_{2ab}+2\xi_{2ab}+2\xi_{3ab})-\frac{32}{27}\Theta^{3}(3\psi_{1ab}+2\xi_{1ab}).\nonumber\\\ear It is straight forward to show that the propagation of $\mathcal{B}_{ab}$ yields,
\be \label{aha}\dot{\mathcal{B}}_{ab}+\sfrac{17}{6}\Theta \mathcal{B}_{ab}-2\mu \mathcal{A}_{ab}=0.\ee The solution to the coupled system of equations given by (\ref{alta}, \ref{Avar}, \ref{aha}) reproduces (\ref{SolF}).
\def\baselinestretch{1}
\section{\label{appb}Commutation Relations}
\def\baselinestretch{1.66}The following
commutation relations are satisfied by any scalar, $f$, and  PSTF-tensor quantity, $T_{ab}$, that are defined on a FLRW background.
\ber
\label{app:0}\varepsilon_{abc}\varepsilon^{dec}&=&2!{h^{d}}_{[a}{h^
{e}}_{b]},\\
\label{app:1}\varepsilon_{abc}{T^{b}}_{p}{T^{p}}_{q}V^{cq}&=&-T_{ab}
\varepsilon^{bcd}{T_{c}}^{p}V_{dp},
\\
\label{app:2}(\tilde{\nabla}_{a}f)^{.}&=&\tilde{\nabla}_{a}\dot{f}-\frac{1}{3}\Theta\tilde{\nabla}_{a}f-{\sigma_{a}}^{b}\tilde{\nabla}_{b}f,
\\
\label{app:4}\tilde{\nabla}_{[a}\tilde{\nabla}_{b]}f&=&0,\ear
\ber\label{app:10}(\curl T_{ab})^{.}&=&
\curl(\dot{T})_{ab}-\frac{1}{3}\Theta \curl
T_{ab}+\epsilon_{cd\langle a}[-\sigma^{ec}D_{|e|}{T_{b\rangle}}^{d}+A^{c}{\dot{T}_{b}}^{d}+\frac{1}{3}\Theta A^{c}{T_{b\rangle}}^{d}+3H_{c\langle}{T_{b\rangle}}^{d}],\nn\\\ear \ber \label{app:11}\curl
\curl(T)_{ab}&=&-D^{2}{T}_{ab}+\frac{3}
{2}D_{\langle a}D^{c}T_{b\rangle
c}+(\mu-\frac{1}{3}\Theta^{2})T_{ab}+3T_{c\langle
a}[{E_{b\rangle}}^{c}-\frac{1}{3}\Theta{\sigma_{b\rangle}}^{c}]+\sigma_{cd}T^{cd}\sigma_{ab}\nonumber\\&&-T^{cd}\sigma_{ca}\sigma_{bd}+\sigma^{cd}\sigma_{c(a}T_{b)d}.\ear

\section*{Bibliography}
\bibliographystyle{unsrt}
\bibliography{SogiNotes-1}{}

\begin{thebibliography}{10}

\bibitem{TCM07}
Tsagas~C G, Challinor A, and Maartens R.
\newblock Relativistic cosmology and large-scale structure.
\newblock {\em Phys. Rept. 465:61-147}, 2008.

\bibitem{WMAPNine}
G.~Hinshaw and et~al.
\newblock Nine-year wilkinson microwave anisotropy probe (wmap) observations:
  Cosmological parameter results.
\newblock {\em Ap J. S.}, 208(2):19, 2013.

\bibitem{Bennett}
C.~L. Bennett and et~al.
\newblock Nine-year wilkinson microwave anisotropy probe (wmap) observations:
  Final maps and results.
\newblock {\em Ap J. S.}, 208(2), 2013.

\bibitem{KODSA84}
H.~Kodama and M.~Sasaki.
\newblock Cosmological perturbation theory.
\newblock {\em Prog.Theor. Phys. Supp.}, 78:213--313, 1984.

\bibitem{MUK92}
V.F. Mukhanov, H.A.Feldman, and R.H. Brandenberger.
\newblock Theory of cosmological perturbations.
\newblock {\em Physics Reports}, 215 (5):203--333, 1992.

\bibitem{Bardeen:1980kt}
J.~M. Bardeen.
\newblock Gauge invariant cosmological perturbations.
\newblock {\em Phys. Rev.}, D22:1882--1905, 1980.

\bibitem{DUR04}
R.~Durrer.
\newblock Cosmological perturbation theory.
\newblock {\em Astro-ph/0402129v2}, 2004.

\bibitem{MMB98}
S.~Mollerach S.~Matarrese and M.~Bruni.
\newblock Relativistic second-order perturbations of the einstein--de sitter
  universe.
\newblock {\em Phys. Rev. D}, 58:043504, 1998.

\bibitem{Nakamura:2006rk}
K.~Nakamura.
\newblock Gauge-invariant formulation of the second-order cosmological
  perturbations.
\newblock {\em Phys. Rev. D}, 74:101301, 2006.

\bibitem{Clarkson:2003af}
C.~A. Clarkson.
\newblock Density fluctuations and gravity waves: A covariant approach to
  gauge-invariant non-linear cosmological perturbation theory.
\newblock {\em Phys. Rev.}, D70:103524, 2004.

\bibitem{Acquaviva:2002ud}
{V. Acquaviva, N. Bartolo, S. Matarrese} and A.~Riotto.
\newblock Second-order cosmological perturbations from inflation.
\newblock {\em Nucl. Phys.}, B667:119--148, 2003.

\bibitem{Brand07}
P.~Martineau and R.~Brandenberger.
\newblock A back-reaction induced lower bound on the tensor-to-scalar ratio.
\newblock {\em Mod. Phys. Lett.}, A23:727--735, 2008.

\bibitem{Osano:2006ew}
{B. Osano, C. Pitrou, P. K. S. Dunsby, J-P. Uzan} and C.~Clarkson.
\newblock Gravitational waves generated by second order effects during
  inflation.
\newblock {\em JCAP}, 0704:003, 2007.

\bibitem{Teresa:2006af}
{T. Lu and A. Kishore} and C.~Clarkson.
\newblock Vector perturbation.
\newblock {\em arXiv:0709.1619}, 2007.

\bibitem{Mena:2002wq}
{F.C. Mena, R. Tavakol} and M.~Bruni.
\newblock Second order perturbations of flat dust flrw universes with a
  cosmological constant.
\newblock {\em Int. J. Mod. Phys.}, A17:4239--4244, 2002.

\bibitem{Matarrese:1997ay}
{S. Matarrese, S. Mollerach} and M.~Bruni.
\newblock Second-order perturbations of the einstein-de sitter universe.
\newblock {\em Phys. Rev.}, D58:043504, 1998.

\bibitem{Finelli:2006wk}
{F. Finelli, G. Marozzi, G.P. Vacca} and G.~G.~Venturi.
\newblock Second order gauge-invariant perturbations during inflation.
\newblock {\em Phys. Rev.}, D74:083522, 2006.

\bibitem{Noh:2004bc}
H.~Noh and J-C. Hwang.
\newblock Second-order perturbations of the friedmann world model.
\newblock {\em Phys. Rev.}, D69:104011, 2004.

\bibitem{Langlois:2005qp}
D.~Langlois and F.~Vernizzi.
\newblock Conserved non-linear quantities in cosmology.
\newblock {\em Phys. Rev.}, D72:103501, 2005.

\bibitem{Langlois:2005ii}
D.~Langlois and F.~Vernizzi.
\newblock Evolution of non-linear cosmological perturbations.
\newblock {\em Phys. Rev. Lett.}, 95:091303, 2005.

\bibitem{Ananda:2006af}
{K. Ananda, and C. Clarkson} and D.~Wands.
\newblock The cosmological gravitational wave background from primordial
  density perturbations.
\newblock {\em Phys. Rev.}, D75:123518, 2007.

\bibitem{Malik2003mv}
K.~A. Malik and D.~Wands.
\newblock Evolution of second order cosmological perturbations.
\newblock {\em Class. Quant. Grav.}, 21:L65--L72, 2004.

\bibitem{Tomita67}
K.~Tomita.
\newblock Non-linear theory of gravitational instability in the expanding
  universe.
\newblock {\em Prog. Theor. Phys.}, 37:831, 1967.

\bibitem{PC96}
T.~Pyne and S.~Carroll.
\newblock Higher-order gravitational perturbations of the cosmic microwave
  background.
\newblock {\em Phys. Rev. D}, 53(2920), 1996.

\bibitem{MM97}
S.~Mollerach and S.~Matarrese.
\newblock Cosmic microwave background anisotropies from second order
  gravitational perturbations.
\newblock {\em Phys. Rev. D}, 56(4480), 1997.

\bibitem{BCLM04}
N.~Bartolo et~al.
\newblock Perturbations in cosmologies with a scalar field and a perfect fluid.
\newblock {\em Phys. Rev}, D70:043532, 2004.

\bibitem{Bartolo:2006cu}
{N. Bartolo, S. Matarrese} and A.~Riotto.
\newblock Cmb anisotropies at second order i.
\newblock {\em JCAP}, 0606:024, 2006.

\bibitem{Bartolo:2006fj}
{N. Bartolo, S. Matarrese} and A.~Riotto.
\newblock Cmb anisotropies at second-order ii: Analytical approach.
\newblock {\em JCAP}, 0701:019, 2007.

\bibitem{Mollerach:2003nq}
{S. Mollerach, D. Harari} and S.~Matarrese.
\newblock Cmb polarization from secondary vector and tensor modes.
\newblock {\em Phys. Rev.}, D69:063002, 2004.

\bibitem{BMMR}
S.~Mollerach N.~Bartolo, S.~Matarrese and A.~Riotto.
\newblock B-mode polarization of the cmb from the second-order photon
  quadrupole.
\newblock {\em arXiv:astro-ph/0703386}, 2007.

\bibitem{EMM}
R.~Maartens G.~F. R.~Ellis and M.~MacCallum.
\newblock {\em Relativistic Cosmology}.
\newblock Number 978-0-521-38115-4. Cambridge University Press, 2012.

\bibitem{EB89}
G.F.R. Ellis and M.~Bruni.
\newblock Covariant and gauge-invariant approach to cosmological density
  fluctuations.
\newblock {\em Phys. Rev. D}, 40:1804, 1989.

\bibitem{Stewart:1974uz}
J.~M. Stewart and M.~Walker.
\newblock Perturbations of spacetimes in general relativity.
\newblock {\em Proc. Roy. Soc. Lond.}, A341:49--74, 1974.

\bibitem{Bruni:1992dg}
{M. Bruni, and P.K.S. Dunsby} and G.F.R. Ellis.
\newblock Cosmological perturbations and the physical meaning of gauge
  invariant variables.
\newblock {\em Astrophys. J.}, 395:34--53, 1992.

\bibitem{Buchert2015}
T.~Buchert {\it et al}.
\newblock Is there proof that back-reaction of inhomogeneities is irrelevant in
  cosmology?
\newblock {\em Class. Quantum Grav.}, 32:44, 2015.

\bibitem{Green2014}
S.~R. Green and R.~M. Wald.
\newblock How well is our universe described by an flrw model?
\newblock {\em Class. Quantum Grav.}, 31:234003, 2014.

\bibitem{EHvE98}
G.F.R. Ellis and H.~van Elst.
\newblock {\em Theoretical and observational Cosmology}.
\newblock edited by Lachieze-Rey, NATO Science series, (Kluwer Academic
  Publishers, Dorretcht), 1999.

\bibitem{EllisCar}
G.~F.~R. Ellis.
\newblock {\em Carge`se Lectures in Physics, Vol. 6, Ed. E Schatzman}.
\newblock Gordon and Breach, New York, 1973.

\bibitem{Bruni:1996im}
{M. Bruni, S. Matarrese, S. Mollerach} and S.~Sonego.
\newblock Perturbations of spacetime: Gauge transformations and gauge
  invariance at second order and beyond.
\newblock {\em Class. Quant. Grav.}, 14:2585--2606, 1997.

\bibitem{CB2011}
C.~Clarkson and B.~Osano.
\newblock Locally extracting scalar, vector and tensor modes in cosmological
  perturbation theory.
\newblock {\em Classical and Quantum Gravity}, 28 (22):225002, 2011.

\bibitem{Clarkson:2011td}
C.~Clarkson and B.~Osano.
\newblock {Locally extracting scalar, vector and tensor modes in cosmological
  perturbation theory}.
\newblock {\em Class.Quantum Grav.}, 28:225002, 2011.

\bibitem{Maartens:1996ch}
{R. Maartens and G. F. R.Ellis} and T.C.S. Siklos.
\newblock Local freedom in the gravitational field.
\newblock {\em Class. Quant. Grav.}, 14:1927--1936, 1997.

\bibitem{RevModPhys.39.862}
E.~R. Harrison.
\newblock Normal modes of vibrations of the universe.
\newblock {\em Rev. Mod. Phys.}, 39(4):862--882, Oct 1967.

\bibitem{Challinor:1999xz}
A.~Challinor.
\newblock Microwave background anisotropies from gravitational waves: The 1+3
  covariant approach.
\newblock {\em Class. Quant. Grav.}, 17:871--889, 2000.

\bibitem{Bond1}
J.~R. Bond and G.~Efstathiou.
\newblock Cosmic background radiation anisotropies in universes dominated by
  non-baryonic dark matter.
\newblock {\em Astrophys. J. Lett}, page 285, 1984.

\bibitem{Bond2}
J.~R. Bond and G.~Efstathiou.
\newblock The statistics of cosmic background radiation fluctuations.
\newblock {\em Mon. Not. R. Astron. Soc}, page 226, 1987.

\bibitem{Zald}
M.~Zaldarriaga and D.~Harari.
\newblock Analytical approach to the polarization of the cosmic microwave
  background in flat and open universes.
\newblock {\em Phys. Rev. D}, 52:3276, 1995.

\bibitem{Plon}
A.~G. Plonarev.
\newblock Polarization and anisotropy induced in the microwave background by
  cosmological gravitational waves.
\newblock {\em Sov. Astron.}, 29:607, 1985.

\bibitem{Critten}
R.~L.~Davis R.~Crittenden and P.~J. Steinhardt.
\newblock Polarization of the microwave background due to primordial
  gravitational waves.
\newblock {\em Astrophys. J. Lett.}, L13:417, 1993.

\bibitem{Harari}
D.~Harari and M.~Zaldarriaga.
\newblock Polarization of the cosmic background in the inflationary cosmology.
\newblock {\em Phys. Lett. B}, 319:96, 1993.

\bibitem{Koswosky}
A.~Koswosky.
\newblock Cosmic microwave background polarization.
\newblock {\em Ann. Phys. (N.Y.)}, 49:246, 1996.

\bibitem{Mel}
A.~Melchiorri and N.~Vittorio.
\newblock Polarization of the microwave background: theoretical framework.
\newblock {\em Report No. astro-phy 9610029.}, page 246, 1996.

\bibitem{Kam}
M.~Kamionkowski et~all.
\newblock A probe of primordial gravity waves and vorticity.
\newblock {\em PRL}, 78:11, 1997.

\bibitem{Gebb}
P.~K. S.~Dunsby T.~Gebbie and G.~F.~R. Ellis.
\newblock 1+3 covariant cosmic microwave background anisotropies ii: The almost
  -friedmann lemaite model.
\newblock {\em Ann. Phys.}, 282, 2000.

\bibitem{bob2017}
B.~Osano.
\newblock Cmb polarization and higher order perturbations.
\newblock {\em In preparation}, 2017.

\end{thebibliography}

\end{document}